\documentclass[aps,prb,twocolumn,reprint,floatfix,nofootinbib]{revtex4-2}
\usepackage{graphicx}
\usepackage{amsmath,amssymb,mathrsfs,amsthm}
\usepackage[pdftex, colorlinks=true, linkcolor=blue, citecolor=blue, urlcolor=blue]{hyperref}
\usepackage{bbold}
\usepackage{times}
\usepackage{mathtools}
\usepackage{braket}
\usepackage{lipsum}

\usepackage{units}

\usepackage{svg}
\usepackage{comment}
\usepackage[normalem]{ulem}
\usepackage{color}
\newcommand{\stkout}[1]{\ifmmode\text{\sout{\ensuremath{#1}}}\else\sout{#1}\fi}

\begin{document}

\title{Multiple polaritonic edge states in a Su-Schrieffer-Heeger chain\\
strongly coupled to a multimode cavity}

\author{Thomas F.\ Allard}
\affiliation{Universit\'e de Strasbourg, CNRS, Institut de Physique et Chimie des Mat\'eriaux de Strasbourg, UMR 7504, F-67000 Strasbourg, France}
\author{Guillaume Weick}
\affiliation{Universit\'e de Strasbourg, CNRS, Institut de Physique et Chimie des Mat\'eriaux de Strasbourg, UMR 7504, F-67000 Strasbourg, France}

\begin{abstract}

A dimerized chain of dipolar emitters strongly coupled to a multimode optical waveguide cavity is studied.
By integrating out the photonic degrees of freedom of the cavity, the system is recast in a two-band model with an effective coupling, so that it mimics a variation of the paradigmatic Su-Schrieffer-Heeger model, which features a nontrivial topological phase and hosts topological edge states. In the strong-coupling regime, the cavity photons hybridize the bright dipolar bulk band into a polaritonic one, renormalizing the eigenspectrum and strongly breaking chiral symmetry.
This leads to a formal loss of the in-gap edge states present in the topological phase while  they merge into the polaritonic bulk band.
Interestingly, however, we find that bulk polaritons entering in resonance with the edge states inherit part of their localization properties, so that multiple polaritonic edge states are observed.
Although these states are not fully localized on the edges, they present unusual properties.
In particular, due to their delocalized bulk part, owing from their polaritonic nature, such edge states exhibit efficient edge-to-edge transport characteristics.
Instead of being degenerate, they occupy a large portion of the spectrum, allowing one to probe them in a wide driving frequency range.
Moreover, being reminiscent of symmetry-protected topological edge states, they feature a strong tolerance to positional disorder.

\end{abstract}

\maketitle

%%%%%%%%%%%%%%%%%%%%%%%%%%%%%%%%%%%%%%%%%%%%%%%%%%%%%%%%%%%%
\section{Introduction}
\label{sec:Introduction}
%%%%%%%%%%%%%%%%%%%%%%%%%%%%%%%%%%%%%%%%%%%%%%%%%%%%%%%%%%%%

A key challenge in the growing field of topological photonics \cite{Ozawa2019,Rider2019,Rider2022} is to understand the interplay between the physics of topological phases of matter \cite{HasanReview2010,Xiao-LiangReview2011}, such as the remarkable presence of topological edge states robust against perturbations \cite{Asboth2016}, and the one of strong light-matter coupling, which has been shown to significantly modify material properties \cite{EbbesenReview2021}.
The underlying effect of the strong light-matter interaction is an effective long-distance coupling mediated by cavity photons, which has proven to be of major importance for topological phenomena \cite{Appugliese2022}.
In particular, an active literature has recently been devoted to extensions of one-dimensional topological models, such as the renowned Su-Schrieffer-Heeger (SSH) model \cite{Su1979} with additional couplings \cite{DiLiberto2014,Wang2018b,Longhi2018,Downing2018,PerezGonzalez2019,Pocock2019,Bello2019,Downing2019,Malki2019,Nie2020,Ling2020,Hsu2020,Nie2021,Jiao2021,Dias2022,PerezGonzalez2022,McDonnell2022,Wei2022,Dmytruk2022,Buendia2023,Pirmoradian2023,Kvande2023}.

Here, we go one step further by addressing the effects of the strong coupling between a multimode optical waveguide and a bipartite chain of emitters (which we consider as ideal, classical dipoles) through a microscopically derived dispersive and spatially dependent light-matter coupling.
Importantly, the consideration of multiple photonic modes, although not often envisaged, has proven to be essential to correctly model cavity-induced effects \cite{Tichauer2021,Ribeiro2022,Allard2022,Tichauer2023}, and is here a key ingredient of our model.
The point dipoles which we consider represent a large variety of physical systems whose main coupling mechanism is dipolar in nature, and governed by classical electromagnetism. 
Such generic emitters model experimental platforms as diverse as subwavelength plasmonic, dielectric, or SiC nanoparticles \cite{Slobozhanyuk2015,Wang2018b,Zhang2020,Allard2021}, magnonic microspheres \cite{Pirmoradian2018,Rameshti2022,Pirmoradian2023}, microwave antennas \cite{Mann2018,Mann2020}, semiconductor excitons \cite{Yuen-Zhou2016}, cold atoms \cite{Browaeys2016,Perczel2017,Wang2018a}, or any other two-level emitters, as they behave as classical dipoles in the single excitation manifold \cite{Asenjo-Garcia2019}.

While a preliminary investigation of such a polaritonic SSH model has been conducted in Ref.~\cite{Downing2019}, highlighting the impact of the light-matter coupling on the topological phases of the system, here, we focus on the fate of the edge states exhibited by the system.
Importantly, to do so, we refine the model derived in Ref.~\cite{Downing2019} in order to avoid any boundary effects that could influence the edge states.
From the hybridization of the dipolar and cavity photon excitations into polaritons, we observe in the strong light-matter coupling regime the formal loss of the in-gap edge states that are present in the topological phase of the original SSH model, with their merging into the polaritonic bulk band.
Although this may at first appear detrimental to the topological properties of the system, here we demonstrate that, interestingly, bulk polaritons in resonance with the formally lost edge states inherit a large edge localization, so that we coin these new cavity-induced states ``polaritonic edge states."

Originating from the diffusion of edge localization onto numerous bulk polaritons, we show that such exotic edge states present properties that are of particular interest.
Specifically, dissipative transport simulations allow us to unveil exceptional polaritonic edge state transport, as well as a wide frequency range at which the latter states can be driven.
Furthermore, the consideration of a disordered bipartite chain enables us to reveal the remarkable tolerance of the polaritonic edge states to positional disorder.

\begin{figure*}[tb]
    \includegraphics[width=1.35\columnwidth]{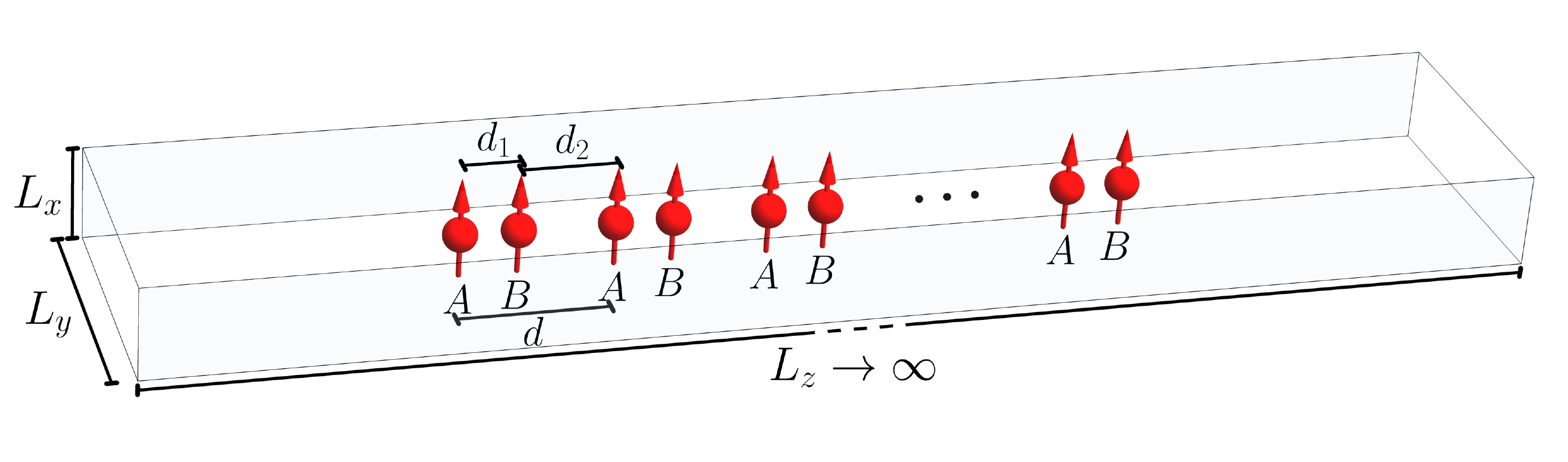}
    \caption{
    Sketch of a dimerized chain of emitters (considered as oscillating dipoles) polarized along the $x$ direction, arranged along the $z$ axis, and placed in the middle of a mirror waveguide cavity with open ends and lengths $L_x$, $L_y$, and $L_z\to \infty$.
    The dipoles, each with bare resonance frequency $\omega_0$,
    belong either to the $A$ or $B$ sublattice and are separated by the alternating 
    distances $d_1$ and $d_2$, so that the lattice constant $d=d_1+d_2$.
    }
\label{fig: sketch}
\end{figure*}

The paper is organized as follows: In Sec.~\ref{sec:Model}, we introduce the model which we use to describe a dimerized chain of emitters strongly coupled to a multimode optical cavity, and derive an effective bipartite Hamiltonian taking into account the effective coupling between the dipoles mediated by the cavity photons.
We study the bulk spectrum of the related two-band Hamiltonian in Sec.~\ref{sec:Bulk spectrum} and 
investigate its topological phases in Sec.~\ref{sec:Topological phases}.
Section \ref{sec:Edge states} is dedicated to the study of the unusual multiple polaritonic edge states that are present in the Hamiltonian finite spectrum, while in Secs.~\ref{sec:Transport} and \ref{sec:Disorder} we study, respectively, their transport properties and their robustness against positional disorder.
Finally in Sec.~\ref{sec:Conclusions}, we summarize our results and discuss further perspectives of our work. Three appendixes 
complement and detail some of the technical aspects of our paper.

%%%%%%%%%%%%%%%%%%%%%%%%%%%%%%%%%%%%%%%%%%%%%%%%%%%%%%%%%%%%%%%%
\section{Dimerized chain coupled to a multimode optical cavity}
\label{sec:Model}
%%%%%%%%%%%%%%%%%%%%%%%%%%%%%%%%%%%%%%%%%%%%%%%%%%%%%%%%%%%%%%%%

We consider a dimerized chain of dipoles coupled to a multimode optical waveguide cavity with perfectly conducting mirrors in the $x$ and $y$ planes, as sketched in Fig.~\ref{fig: sketch}.
To model such a system, we rely on the approach developed in Ref.~\cite{Downing2019} and employ the polaritonic Hamiltonian within the Coulomb gauge
\begin{equation}
    H = H_{\mathrm{dp}} + H_{\mathrm{ph}} + H_{\mathrm{dp}\textrm{-}\mathrm{ph}}.
\label{eq:Hamiltonian}
\end{equation}

%===================================================
\subsection{Quasistatic dipolar Hamiltonian}
\label{sec:Model A}

The first term on the right-hand side (r.h.s.)\ of Eq.~\eqref{eq:Hamiltonian} corresponds to the Hamiltonian of a finite dimerized chain of $2\mathcal{N}$ dipoles, where $\mathcal{N}$ is the number of unit cells,\footnote{We consider a lattice with an even number of sites.} coupled through a quasistatic dipole-dipole interaction.
Ignoring counter-rotating terms, such dipolar Hamiltonian reads \cite{Downing2018}
\begin{align}
    H_{\mathrm{dp}} =&\; \hbar\omega_0 \sum_{m=1}^\mathcal{N} 
    \left(a_{m}^{\dagger} a_{m}^{\phantom{\dagger}}+b_{m}^{\dagger} b_{m}^{\phantom{\dagger}}\right)
    \nonumber \\ 
    &+ \frac{\hbar\Omega}{2}\sum_{\substack{m,m'=1\\(m\neq m')}}^{\mathcal{N}} 
    f_{m-m'}\left( a_{m^{\phantom{\prime}}}^{\dagger} a_{m^\prime}^{\phantom{\dagger}} + b_{m^{\phantom{\prime}}}^{\dagger} b_{m^\prime}^{\phantom{\dagger}} + \mathrm{H.c.} \right)
    \nonumber \\
    &+ \hbar\Omega\sum_{m,m'=1}^{\mathcal{N}} g_{m-m'} \left( a_{m^{\phantom{\prime}}}^{\dagger} b_{m^\prime}^{\phantom{\dagger}} + \mathrm{H.c.} \right).
\label{eq:H_dp real}
\end{align}
In the expression above, the bosonic operators $a_m^\dagger$ ($b_m^{\dagger}$) and $a_m^{\phantom{\dagger}}$ ($b_m^{\phantom{\dagger}}$) create and annihilate, respectively, a dipolar excitation on the site $m$ of the sublattice $A$ ($B$), polarized along the $x$ axis and with resonance frequency $\omega_0$.
The strength of the all-to-all quasistatic dipolar coupling is $\Omega=(\omega_0/2)(a/d)^3$, with $d$ the lattice constant of the chain, and where $a=(Q^2/M\omega_0^2)^{1/3}$ is a lengthscale which characterizes the strength of each point dipole with charge $Q$ and mass $M$.
The Hamiltonian \eqref{eq:H_dp real} represents a variation of the dipolar SSH model of Ref.~\cite{Downing2017_Topological}, which, as a toy model, only considered nearest-neighbor coupling.
Here, the intra- ($A\!\leftrightarrow\!A$ or $B\!\leftrightarrow\!B$) and intersublattice ($A\!\leftrightarrow\!B$) all-to-all quasistatic couplings are included and read, respectively \cite{Downing2018},
\begin{subequations}
\label{eq:lattice sums real}
    \begin{equation}
        f_{m-m'} = \frac{1}{|m-m'|^3}
    \label{eq:f_nm}
    \end{equation}
and
    \begin{equation}
        g_{m-m'} = \frac{ 1 }{ |m - m' - d_{1}/d |^3 },
    \label{eq:g_nm}
    \end{equation}
\end{subequations}
where $d_1$ is the intradimer distance.
The difference between the dipole-dipole distances $d_1$ and $d_2$ defined in Fig.~\ref{fig: sketch} encodes the dimerization of the chain, and is quantified by the dimerization parameter 
\begin{equation}
\label{eq:epsilon_dimerization}
    \epsilon=\frac{d_1-d_2}{d}.
\end{equation}

While the intersublattice coupling \eqref{eq:g_nm} preserves the chiral symmetry of the model, the intrasublattice one \eqref{eq:f_nm} breaks it, destroying a priori the (chirally-protected) topological phase of the SSH model.\footnote{In the language of the ten-fold way \cite{Ryu_2010}, the latter intrasublattice coupling leads the Hamiltonian \eqref{eq:H_dp real} to belong to the AI symmetry class, which is trivial in one dimension, instead of belonging to the AIII symmetry class of the so-called generalized SSH model with all-neighbor intersublattice coupling \cite{Chen2020}, which has a $\mathbb{Z}$ topological invariant.}
However, inversion symmetry still allows the dipolar Hamiltonian \eqref{eq:H_dp real} to feature a quantized Zak phase as well as edge states \cite{Hughes2011,Miert2016}.

The model encapsulated in Eq.~\eqref{eq:H_dp real} has been studied in Ref.~\cite{Downing2018}, where it has been shown that due to the fast, $1/|m-m'|^3$ decrease and the small value of the quasistatic intrasublattice dipolar coupling, the spectrum of the Hamiltonian, as well as its topological phases, are almost unchanged from the original SSH model.
Its edge states, however, are naturally not anymore protected by chiral symmetry, and hence do not show a formal robustness against chiral-preserving disorder.

%===================================================
\subsection{Coupling to a multimode optical waveguide cavity}
\label{sec:Model B}

The second and third terms on the r.h.s.\ of Eq.~\eqref{eq:Hamiltonian} account, respectively, for the photonic degrees of freedom and their minimal coupling to the dipolar Hamiltonian presented above.

In contrast to Ref.~\cite{Downing2019}, where hard wall boundary conditions for the cavity in the three space directions have been considered when studying the finite system, here we consider periodic boundary conditions in the $z$ direction and perfect mirrors in the $x$ and $y$ directions only, so that we model a finite dipole chain embedded into an infinite cavity with longitudinal size $L_z \to \infty$ (see Fig.~\ref{fig: sketch}).
This limit is equivalent to an open waveguide cavity, with open boundary conditions in the $z$ direction.
Our motivation to consider such waveguide cavity is twofold.
First, it represents a more feasible experimental realization than a closed cuboidal cavity.
Second, as we are in particular interested in the properties of the edge states forming around the first and last dipole of the chain, we wish to avoid any boundary effects due to the cavity walls at the two ends of the chain.
While this choice of boundary condition could be viewed, at first sight, as a minor change, it in fact drastically affects some of the properties of the system, as discussed in Appendix \ref{sec:Appendix A: Strong coupling Hamiltonian}.

By quantizing the electromagnetic field inside the photonic cavity depicted in Fig.~\ref{fig: sketch}, one obtains \cite{Kakazu1994} an infinite number of photonic modes with dispersion 
\begin{equation}
    \omega^{\mathrm{ph}}_{\mathbf{k},l}=c|\mathbf{k}-\mathbf{G}_l|.
\label{eq:dispersion photons}
\end{equation}
Here, $c$ is the speed of light in vacuum, and $\mathbf{k}=(\pi n_x/L_x, \pi n_y/L_y, q)$ is the photonic wavevector, with $n_x,n_y\in\mathbb{N}$, while the longitudinal quasimomentum $q\in[-\pi/d,+\pi/d]$ belongs to the first Brillouin zone.
$\mathbf{G}_l=2\pi l \hat{z}/d$ ($l \in \mathbb{Z}$) represents the set of reciprocal lattice vectors, 
so that the dispersion \eqref{eq:dispersion photons} is $2\pi/d$ periodic.
The consideration of all these photonic bands that are folded within the first Brillouin zone, known as (photon) Umklapp processes or diffraction orders, is here justified by our wish to compute bulk topological quantities in Sec.~\ref{sec:Topological phases}, which formally require periodicity in the first Brillouin zone.

As justified in detail in Ref.~\cite{Downing2019}, by choosing a specific cavity geometry $L_y = 3L_x$ and $3a \lesssim L_x \lesssim 20a$, only the lowest photonic band $(n_x,n_y,q)=(0,1,q)$ is at resonance with the two dipolar ones.
This allows us to consider solely the photonic modes with dispersion
\begin{equation}
   \omega^{\mathrm{ph}}_{q,l} = c\,\sqrt{ \left( \frac{\pi}{L_y} \right)^2 + q_l^2}
   %\left(q - \frac{2\pi l}{d}\right)^2 }\,,
\label{eq:photonic dispersion pbc}
\end{equation}
with $q_l=q-2\pi l/d$, 
so that the photonic Hamiltonian in Eq.~\eqref{eq:Hamiltonian} reads
\begin{equation}
    H_\mathrm{ph} = \sum_{q,l} \hbar\omega^{\mathrm{ph}}_{q,l}{c^{\dagger}_{q,l}} c^{\phantom{\dagger}}_{q,l}.
\label{eq:H_ph pbc}
\end{equation}
Here, the bosonic operator $c_{q,l}^{\dagger}$ ($c_{q,l}^{\phantom\dagger}$) creates (annihilates) a transverse photon with longitudinal quasimomentum $q$ and Umklapp band index $l$.

As discussed in Appendix \ref{sec:Appendix A: Strong coupling Hamiltonian}, the light-matter, minimal coupling Hamiltonian in Eq.~\eqref{eq:Hamiltonian} then reads~\cite{Downing2019}
\begin{align}
   H_{\mathrm{dp}\textrm{-}\mathrm{ph}} =&\; \mathrm{i}\hbar\sum_{m=1}^{\mathcal{N}}\sqrt{\frac{d}{L_z}} \sum_{q,l} \xi_{q,l} \,\mathrm{e}^{\mathrm{i}mq_ld}  \nonumber \\
   &\times \left(  a_m^\dagger\,\mathrm{e}^{-\mathrm{i}\chi_{q,l}} + b_m^\dagger\, \mathrm{e}^{\mathrm{i}\chi_{q,l}}  \right)c_{q,l}^{\phantom\dagger} + \mathrm{H.c.},
\label{eq:H_dp-ph pbc}
\end{align}
with the light-matter coupling strength
\begin{equation}
    \xi_{q,l} = \omega_0\,\sqrt{ \frac{2\pi a^3 \omega_0}{dL_xL_y\omega_{q,l}^\mathrm{ph}}  }\,.
\label{eq:light-matter coupling strength pbc}
\end{equation}
The phase $\chi_{q,l}=q_ld_1/2$ in Eq.~\eqref{eq:H_dp-ph pbc} encodes the fact that photons interact differently with dipolar excitations belonging to the $A$ or $B$ sublattice, due to the spatial dependence of the light-matter coupling inside the cavity.

To conclude the presentation of the model, we note that the effect of image dipoles originating from the cavity walls, as well as counter-rotating terms, are neglected in our description.
We have checked that such additional terms, just as including higher photonic bands, do not qualitatively change any of the results of our paper \cite{Allard2023_PhD}.

%===================================================
\subsection{Effective dipolar Hamiltonian}
\label{sec:Model C}

Since we are mainly concerned by how the strong light-matter interaction renormalizes the dipolar subsystem, we integrate out the photonic degrees of freedom by performing a Schrieffer-Wolff transformation \cite{Schrieffer1966} on the Hamiltonian \eqref{eq:Hamiltonian}.
Such a unitary transformation, which we detail in Appendix \ref{sec:Appendix B: Schrieffer-Wolff transformation}, allows us to perturbatively decouple the photonic and dipolar subspaces to second order in the light-matter coupling strength \eqref{eq:light-matter coupling strength pbc}.

Focusing on the dipolar subspace, we obtain the effective bipartite Hamiltonian
\begin{align}
    \tilde{H}_{\mathrm{dp}} =&\; \hbar\tilde{\omega}_0 \sum_{m=1}^\mathcal{N} 
    \left(a_{m}^{\dagger} a_{m}^{\phantom{\dagger}}+b_{m}^{\dagger} b_{m}^{\phantom{\dagger}}\right)
    \nonumber \\ 
    &+ \frac{\hbar\Omega}{2}\sum_{\substack{m,m'=1\\(m\neq m')}}^{\mathcal{N}} \tilde{f}_{m-m'}\left( a_{m^{\phantom{\prime}}}^{\dagger} a_{m^\prime}^{\phantom{\dagger}} + b_{m^{\phantom{\prime}}}^{\dagger} b_{m^\prime}^{\phantom{\dagger}} + \mathrm{H.c.} \right)
    \nonumber \\
    &+ \hbar\Omega\sum_{m,m'=1}^{\mathcal{N}} \tilde{g}_{m-m'} \left( a_{m^{\phantom{\prime}}}^{\dagger} b_{m^\prime}^{\phantom{\dagger}} + \mathrm{H.c.} \right).
\label{eq:H_dp SW real}
\end{align}
Here, the onsite frequency $\omega_0$ and the intra- and intersublattice sums 
$f_{m-m'}$ and $g_{m-m'}$ are renormalized by the cavity photons [compare with Eq.~\eqref{eq:H_dp real}] as
\begin{subequations}
\label{eq: SW renormalized quantities real}
    \begin{equation}
        \tilde{\omega}_0 = \omega_0 - \frac{d}{2\pi}\sum_{l=-\infty}^{+\infty}\int_{-\pi/d}^{+\pi/d} \mathrm{d}q\, \frac{\xi^2_{q,l}}{\omega_{q,l}^{\mathrm{ph}} - \omega_0},
    \label{eq: SW renormalized omega0}
    \end{equation}
    \begin{align}
        \tilde{f}_{m-m'} =&\; f_{m-m'} \nonumber \\
        &- \frac{1}{\Omega}\frac{d}{2\pi}\sum_{l=-\infty}^{+\infty}\int_{-\pi/d}^{+\pi/d} \mathrm{d}q\, \frac{\xi^2_{q,l}\, \mathrm{e}^{\mathrm{i}(m-m')q_ld}}{\omega_{q,l}^{\mathrm{ph}} - \omega_0},
    \label{eq: SW renormalized intersublattice}
    \end{align}
and
    \begin{align}
        \tilde{g}_{m-m'} =&\; g_{m-m'} \nonumber \\
        &- \frac{1}{\Omega}\frac{d}{2\pi}\sum_{l=-\infty}^{+\infty}\int_{-\pi/d}^{+\pi/d} \mathrm{d}q\, \frac{\xi^2_{q,l}\,\mathrm{e}^{\mathrm{i}(m-m'-d_1/d)q_ld} }{\omega_{q,l}^{\mathrm{ph}} - \omega_0},
    \label{eq: SW renormalized intrasublattice}
    \end{align}
\end{subequations}
where we took the continuous limit for the quasimomentum $q$, by considering an infinitely long cavity with $L_z/d \rightarrow \infty$.

Our real-space perturbation theory provides a transparent interpretation of the effects of the strong light-matter coupling, as the above renormalized quantities account for an effective coupling between the dipoles which is mediated by the cavity photons.
Importantly, we have checked that such perturbation theory qualitatively reproduces the results obtained from a diagonalization of the full polaritonic Hamiltonian \eqref{eq:Hamiltonian}, as long as the cavity height $L_x \lesssim 10a$.

The integrals in Eq.~\eqref{eq: SW renormalized quantities real} are evaluated analytically in Appendix~\ref{sec:Appendix C: Effective coupling integrals}.
On the one hand, the quasistatic, power-law dipole-dipole couplings $f_{m-m'}$ and $g_{m-m'}$ of Eq.~\eqref{eq:lattice sums real} are renormalized by the addition of a quasi-exponential decay, whose characteristic length is governed by the cavity transverse dimensions $L_x$ and $L_y$ [see Eq.~\eqref{eq:effective coupling integral J_2}]. 
On the other hand, the bare dipole frequency $\omega_0$ is only slightly redshifted [see Eq.~\eqref{eq:effective coupling integral I(0)}].

Tuning the cavity transverse dimensions so that the photon dispersion \eqref{eq:photonic dispersion pbc} approaches resonance with the two dipolar bands then allows one to enter in the strong light-matter coupling regime.
Hence, the Hamiltonian \eqref{eq:H_dp SW real} can be viewed as a variation of a dipolar SSH model, with hoppings being highly modified by the strong coupling to a multimode optical cavity.
In the sequel of this paper, we therefore refer to such a model as a polaritonic SSH model.

\begin{figure*}[tb]
    \includegraphics[width=\linewidth]{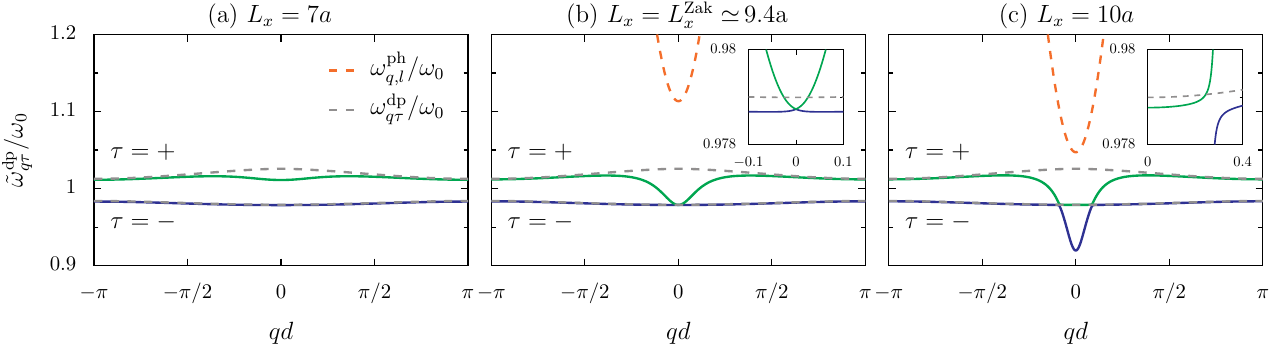}
    \caption{Green and blue solid lines: Polaritonic dispersions $\tilde{\omega}^\mathrm{dp}_{q,\tau=\pm}$ from Eq.~\eqref{eq:dispersion SW}, in units of the bare dipole frequency $\omega_0$ and in the first Brillouin zone 
    for cavity heights (a) $L_x=7a$, (b) $L_x=L_x^\mathrm{Zak}\simeq9.4a$, and (c) $L_x=10a$.
    Gray and orange dashed lines: Bare dipolar dispersion $\omega_q^\mathrm{dp}$ [obtained by replacing tilded by nontilded quantities in Eq.~\eqref{eq:dispersion SW}] and photonic one $\omega_{q,l}^\mathrm{ph}$ [Eq.~\eqref{eq:photonic dispersion pbc}], respectively. Only the lowest, $l=0$, photonic band is visible on the figure.
    Inset of panel (b): Detail of the gap closing at $q=0$ taking place when $L_x=L_x^\mathrm{Zak}$.
    Inset of panel (c): Detail of the avoided crossing between the two dipolar bands appearing from $L_x > L_x^\mathrm{Zak}$.
    In the figure, the dimerization parameter $\epsilon=0.25$ [cf.\ Eq.~\eqref{eq:epsilon_dimerization}], and the Umklapp index $l\in[-l_\mathrm{max}, +l_\mathrm{max}]$ with $l_\mathrm{max}=100$. As in the remaining of the paper, the lattice constant $d=8a$ and the dimensionless dipole strength $k_0a=\omega_0 a/c=0.1$.}
\label{fig:Dispersions SW}
\end{figure*}

Although strong coupling does not fundamentally modify the symmetries of the effective Hamiltonian [compare Eqs.~\eqref{eq:H_dp real} and \eqref{eq:H_dp SW real}], it can significantly increase the chiral-breaking, intrasublattice interaction \eqref{eq: SW renormalized intersublattice}, which can become dominant over the chiral-preserving, intersublattice one \eqref{eq: SW renormalized intrasublattice}.
As we will discuss in the next Sec.~\ref{sec:Bulk spectrum}, this increased asymmetry between the two sublattices strongly modifies the bulk spectrum. As we will show in Sec.~\ref{sec:Topological phases}, the topological phases of the effective Hamiltonian \eqref{eq:H_dp SW real} are also affected by the light-matter coupling.

%%%%%%%%%%%%%%%%%%%%%%%%%%%%%%%%%%%%%%%%%%%%%%%%%%%%%%
\section{Bulk Hamiltonian and Fourier diagonalization}
\label{sec:Bulk spectrum}
%%%%%%%%%%%%%%%%%%%%%%%%%%%%%%%%%%%%%%%%%%%%%%%%%%%%%%

To study the bulk spectrum of the polaritonic SSH model, we consider here the thermodynamic limit where the number of emitters $2\mathcal{N}$ goes to infinity, so that we can use periodic boundary conditions for the chain.

By performing a Fourier transformation on the effective Hamiltonian \eqref{eq:H_dp SW real}, we obtain an effective two-band Hamiltonian in reciprocal space $\tilde{H}_\mathrm{dp} = \sum_q {{\psi}}_q^\dagger \tilde{\mathcal{H}}_q^{\phantom{\dagger}} {{\psi}}_q^{\phantom{\dagger}}$, with the Bloch Hamiltonian
\begin{equation}
    \tilde{\mathcal{H}}_q = \hbar
    \begin{pmatrix}
        \omega_{0} + \Omega \tilde{f}_q & \Omega \tilde{g}_q \\ 
        \Omega \tilde{g}_q^* & \omega_{0} + \Omega \tilde{f}_q
    \end{pmatrix}.
\label{eq:SW Bloch fourier}
\end{equation}
Here, the spinor creation operator $\psi_q^\dagger = (a_q^\dagger, b_q^\dagger)$, where the bosonic ladder operators $a_q^\dagger$ ($b_q^\dagger$) and $a_q^{\phantom\dagger}$ ($b_q^{\phantom\dagger}$) create and annihilate, respectively, a dipolar excitation with resonance frequency $\omega_0$ polarized along the $x$ axis on the $A$ ($B$) sublattice, with longitudinal quasimomentum $q\in[-\pi/d,+\pi/d]$ in the first Brillouin zone.
The cavity-renormalized intra- and intersublattice sums in Eq.~\eqref{eq:SW Bloch fourier} read in Fourier space
\begin{subequations}
\label{eq:lattice sums SW}
\begin{equation}
    \tilde{f}_q = f_q - \frac{1}{\Omega}\, \sum_{l=-\infty}^{+\infty} \frac{\xi^2_{q,l}}{\omega_{q,l}^\mathrm{ph} - \omega_{0}}
    \label{eq:lattice sum f SW}
\end{equation}
and
\begin{equation}
    \tilde{g}_q = g_q - \frac{1}{\Omega} \, \sum_{l=-\infty}^{+\infty} \frac{\xi^2_{q,l}\;\mathrm{e}^{-2\mathrm{i}\chi_{q,l}}}{\omega_{q,l}^\mathrm{ph} - \omega_{0}},
    \label{eq:lattice sum g SW}
\end{equation}
\end{subequations}
respectively,
where the reciprocal counterparts of the bare sublattice sums \eqref{eq:lattice sums real} are \cite{Downing2018}
\begin{subequations}
    \begin{equation}
        f_q = 2\sum_{m=1}^\infty \frac{\cos(mqd)}{m^3}
    \label{eq:lattice sum fq}
    \end{equation}
and
    \begin{equation}
        g_q = \sum_{m=0}^\infty \left[ \frac{\mathrm{e}^{\mathrm{i}mqd}}{(m+d_1/d)^3} + \frac{\mathrm{e}^{-\mathrm{i}(m+1)qd}}{(m+d_2/d)^3} \right].
    \label{eq:lattice sum gq}
    \end{equation}
\label{eq:lattice sums fourier}
\end{subequations}

A Bogoliuobov transformation of the effective Bloch Hamiltonian \eqref{eq:SW Bloch fourier} leads to the eigenfrequencies
\begin{equation}
    \tilde{\omega}^\mathrm{dp}_{q\tau} = {\omega}_{0} + \Omega \tilde{f}_q + \tau\, \Omega |\tilde{g}_q|,
\label{eq:dispersion SW}
\end{equation}
where $\tau=+$ $(-)$ denotes the high- (low-)energy band, and to the eigenspinors 
\begin{equation}
\label{eq:renormalized_spinors}
\ket{ \tilde{\psi}_{q\tau} } = 
\frac{1}{\sqrt{2}}
\begin{pmatrix}
1 \\
\tau\, \mathrm{e}^{\mathrm{i}\tilde{\phi}_q}
\end{pmatrix},
\end{equation}
where the phase $\tilde{\phi}_q = \arg(\tilde{g}_q)$.

The renormalized dipolar dispersion \eqref{eq:dispersion SW} is shown for increasing cavity dimensions [which encode the light-matter coupling strength \eqref{eq:light-matter coupling strength pbc}] in Fig.~\ref{fig:Dispersions SW} in the first Brillouin zone for a dimerization parameter $\epsilon=0.25$ [cf.\ Eq.~\eqref{eq:epsilon_dimerization}]. 
The upper $(\tau=+)$ and lower $(\tau=-)$ bands are displayed by green and blue solid lines, respectively.
In the figure, 
we further plot the bare dipolar dispersion $\omega_q^\mathrm{dp}$ [obtained by replacing tilded by nontilded quantities in Eq.~\eqref{eq:dispersion SW}] and the bare photonic one $\omega_{q,l}^\mathrm{ph}$ [Eq.~\eqref{eq:photonic dispersion pbc}] by gray and orange dashed lines, respectively.

In Fig.~\ref{fig:Dispersions SW}(a), we consider a cavity height $L_x=7a$.
In such a case, the photonic modes [not visible on the scale of Fig.~\ref{fig:Dispersions SW}(a)] are too high in energy to significantly renormalize the dipolar bands.
Notably, only the upper polaritonic effective branch (green solid line) is redshifted around the center of the Brillouin zone, while the lower one (blue solid line) is essentially unaffected by the light-matter coupling.
On the one hand, only the dispersion at the center of the Brillouin zone is renormalized due to the fact that all the modes with large wavenumber are out of resonance with the photons.
On the other hand, the asymmetric behavior between the two dipolar bands can be understood physically from the fact that for the transverse dipole-dipole interaction at play here, the antiparallel alignment of the dipole moments within a dimer ($\uparrow \downarrow$) is energetically favored.
It leads the low-energy band to behave as a dark band, which only weakly couples to light, while the high-energy band, which favors parallel alignment of the dipoles ($\uparrow \uparrow$), can significantly couple to light and is thus referred to as being a bright band.

When increasing the cavity height, the photonic modes become closer in energy from the bare dipolar ones, so that the bright band ($\tau=+$) is further renormalized in a standard avoided-crossing scheme, while the dark one ($\tau=-$) still remains unaffected by the light-matter coupling.
Such growing asymmetry between the two bands makes clear the broken chiral symmetry of the model, boosted by the cavity-induced renormalization of the intersublattice sum $\tilde{f}_q$ in Eq.~\eqref{eq:lattice sum f SW}. 
This allows the bright modes around the center of the Brillouin zone to increasingly fill the gap between the two bands.
At a cavity height $L_x^\mathrm{edge}$ (not shown), whose significance will become clearer in the next sections, approximately half of the gap is filled.
Through our effective two-band model, we find the latter cavity height to be close to
\begin{equation}
    \frac{L_x^\mathrm{edge}}{a} \simeq \frac{\pi}{3k_0a} - \frac{8k_0a}{f_0 + g_0 + 0.002\omega_0/\Omega} \left(\frac{d}{a}\right)^2 , 
\label{eq:Lxedge}
\end{equation}
where we approximated the middle of the gap 
to $0.998\omega_0$.
In the above equation, $k_0=\omega_0/c$, and, importantly, $g_0$ depends on the dimerization parameter [see Eq.~\eqref{eq:lattice sum gq}].
With the parameters of Fig.~\ref{fig:Dispersions SW}, 
one has $L_x^\mathrm{edge} \simeq 8.7a$.

At an even larger cavity height, the bright band fills entirely the energy gap (for all $q$'s), so that the system is not anymore in an ``insulating" phase, but becomes ``metallic" in the language of condensed matter electronic systems.
In our effective model, such transition occurs when $\tilde{\omega}^\mathrm{dp}_{q=0, \tau=+}=\tilde{\omega}^\mathrm{dp}_{q=\pi/d,\tau=-}$, at a cavity height coined in Ref.~\cite{Downing2019} as $L_x^\mathrm{gap}$.
Using the same parameters as in Fig.~\ref{fig:Dispersions SW}, $L_x^\mathrm{gap} \simeq 9.3a > L_x^\mathrm{edge}$.

Figure \ref{fig:Dispersions SW}(b) displays our results for the cavity height $L_x=L_x^\mathrm{Zak}\simeq9.4a$, where, importantly, the upper band (green line) 
touches the lower one (blue line) at $q=0$, as highlighted in the inset where a zoom of the two curves around the center of the Brillouin zone is shown.
Such critical cavity height $L_x^\mathrm{Zak}$ has been introduced in Ref.~\cite{Downing2019}, and bears its name from the topological phase transition (TPT) that arises here, and which is associated with a modification of the Zak phase.
Our effective two-band model allows us to easily interpret this result analytically.
Indeed, from Eqs.~\eqref{eq:lattice sums SW}--\eqref{eq:dispersion SW}, we have $\tilde{\omega}^\mathrm{dp}_{q=0, \tau=+}=\tilde{\omega}^\mathrm{dp}_{q=0,\tau=-}$ when the renormalization of the sublattice sum \eqref{eq:lattice sum g SW} due to the light-matter coupling  counteracts the original sublattice sum \eqref{eq:lattice sum gq}, i.e., $|\tilde{g}_{0}|=0$.
Considering only the lowest Umklapp index $l=0$ for simplification, this arises for the cavity height
\begin{equation}
    \frac{L_x^\mathrm{Zak}}{a} =  \frac{\pi}{3k_0 a} - \frac{4k_0a}{g_0}\left(\frac{d}{a}\right)^2.
\label{eq:LxZak}
\end{equation}
A detailed discussion of the unusual topological phases of our system is presented in the next Sec.~\ref{sec:Topological phases}.

In Fig.~\ref{fig:Dispersions SW}(c), we further increase the cavity height to $L_x=10a$, so that the proximity in energy of the photonic modes (orange line) redshifts the upper, bright band (green line) into the lower, dark one (blue line).
This results in another avoided-crossing scheme, now between the two effective upper ($\tau=+$) and lower ($\tau=-$) dipolar bands, as highlighted in the inset of Fig.~\ref{fig:Dispersions SW}(c), where a zoom on the two curves is provided.
Hence, as long as $L_x>L_x^\mathrm{Zak}$ the two effective bands anticross, so that the bandgap for a fixed wavenumber $q$ is open again.
However, we emphasize that, as was already the case in Fig.~\ref{fig:Dispersions SW}(b), the energy gap for all $q$'s is closed, so that the system is here metallic.

%%%%%%%%%%%%%%%%%%%%%%%%%%%%%%%%%%%%%%%%%%%%%%%%%%%%%%%%%%%%
\section{Topological phases}
\label{sec:Topological phases}
%%%%%%%%%%%%%%%%%%%%%%%%%%%%%%%%%%%%%%%%%%%%%%%%%%%%%%%%%%%%

The topological phases of the present model have been partly investigated in Ref.~\cite{Downing2019}, where, specifically, the fate of the bulk-boundary correspondence under strong coupling was examined, that is, the accordance between bulk-related topological invariants and the number of edge states in the finite-size system.
In this section, we further investigate the topological phases of the polaritonic SSH model \eqref{eq:Hamiltonian} and deepen the results of Ref.~\cite{Downing2019}.
In particular, we explore both the $\epsilon<0$ and $\epsilon>0$ cases [see Eq.~\eqref{eq:epsilon_dimerization}] and we gain physical insight on the topology of the system by analyzing it through our simpler, effective two-band model.

Despite its broken chiral symmetry ($\{\tilde{\mathcal{H}}_q,\sigma_z\}\neq0$), the effective two-band Hamiltonian \eqref{eq:SW Bloch fourier} conserves inversion symmetry ($\sigma_x\tilde{\mathcal{H}}_{-q} = \tilde{\mathcal{H}}_q\sigma_x$), where $\sigma_x$ and $\sigma_z$ denote the first and third Pauli matrix, respectively.
Importantly, this ensures that the Zak phase \cite{Delplace2011}
\begin{equation}
    \tilde{\vartheta}^\mathrm{Zak} = \mathrm{i}\int_{-\pi/d}^{+\pi/d}  \mathrm{d}q \braket{\tilde{\psi}_{q\tau} | \partial_q \tilde{\psi}_{q\tau}} \quad \mathrm{mod} \; 2\pi 
\label{eq:Zak phase}
\end{equation}
is quantized, and defines a meaningful $\mathbb{Z}_2$ topological invariant of the model \cite{Hughes2011,Miert2016}.\footnote{We note that, in contrast to the terminology used in most of the literature, Eq.~\eqref{eq:Zak phase} does not formally represent a Zak phase but rather $\pi$ times a winding number defined with respect to a specific choice of unit cell, as discussed in detail in Ref.~\cite{Fuchs2021}.}
Due to inversion symmetry, the Zak phase depends only on the behavior of the system at the inversion-invariant momenta $q=0$ and $q=\pm\pi/d$.
We evaluate Eq.~\eqref{eq:Zak phase} using the Wilson-loop approach \cite{Wang_2019}, which is gauge invariant as well as suitable for numerical implementation.

\begin{figure}[tb]
    \includegraphics[width=\columnwidth]{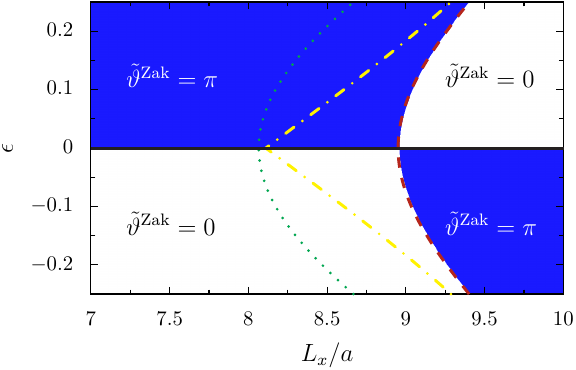}
    \caption{Topological phase diagram from the computation of the Zak phase $\tilde{\vartheta}^\mathrm{Zak}$ [see Eq.~\eqref{eq:Zak phase}] in the $(L_x,\epsilon)$ parameter space.
    The green dotted, yellow dashed-dotted, and red dashed lines correspond, respectively, to $L_x=L_x^\mathrm{edge}$ [see Eq.~\eqref{eq:Lxedge}],  $L_x=L_x^\mathrm{gap}$, and $L_x=L_x^\mathrm{Zak}$ [see Eq.~\eqref{eq:LxZak}]. In the figure, the Umklapp index $l\in[-l_\mathrm{max}, +l_\mathrm{max}]$ with $l_\mathrm{max}=100$.}
\label{fig:Zak phase diagram}
\end{figure}

Our results of such computation of the Zak phase \eqref{eq:Zak phase} are shown in Fig.~\ref{fig:Zak phase diagram} as a phase diagram in the $(L_x,\epsilon)$ parameter space.
Two TPTs between the trivial ($\tilde{\vartheta}^\mathrm{Zak}=0$, white regions) and topological phases ($\tilde{\vartheta}^\mathrm{Zak}=\pi$, blue regions) are visible.
A first one, induced by the variation of the dimerization of the chain, is present at $\epsilon=0$ and indicated as a black solid line.
Such a transition characterizes the two topological phases of the original SSH model, and results from a bandgap closing at $q=\pm\pi/d$.
The second TPT, indicated by a red dashed line, arises at $L_x=L_x^\mathrm{Zak}$ [see Eq.~\eqref{eq:LxZak}], and is solely induced by the light-matter coupling which leads to a bandgap closing at $q=0$ [see Fig.~\ref{fig:Dispersions SW}(b)].

Due to this cavity-induced transition, the nontrivial and trivial phases of the model in the weak-coupling regime (lower and upper left regions in Fig.~\ref{fig:Zak phase diagram}), which, notably, are similar to that of the original SSH model, are reversed in the strong-coupling regime (lower and upper right regions in Fig.~\ref{fig:Zak phase diagram}).

Importantly, the cavity-induced TPT happens once the system is already metallic, since $L_x^\mathrm{Zak}$ (red dashed line in Fig.~\ref{fig:Zak phase diagram}) is larger than $L_x^\mathrm{gap}$ (yellow dashed-dotted line).
Therefore, as will be discussed in the next section, while the dimerization-induced TPT (solid black line) is associated with the presence or absence of edge states as in the original SSH model, the cavity-induced TPT (red dashed line), which separates two gapless metallic phases ($ L_x^\mathrm{gap} < L_x < L_x^\mathrm{Zak}$ and $L_x > L_x^\mathrm{Zak}$), does not influence the presence (or absence) of edge states. 
This led Ref.~\cite{Downing2019} to conclude on the breakdown of the bulk-edge correspondence for this system.\footnote{Note that in an (unphysical) model which would conserve chiral symmetry, i.e., Hamiltonian \eqref{eq:SW Bloch fourier} with $\tilde{f}_q=0$, such cavity-induced TPT would \emph{not} break the bulk-edge correspondence, since $L_x^\mathrm{edge}=L_x^\mathrm{gap}=L_x^\mathrm{Zak}$, so that the system remains in an ``insulating" phase, except at the transition point.}

\begin{figure*}[tb]
    \includegraphics[width=\linewidth]{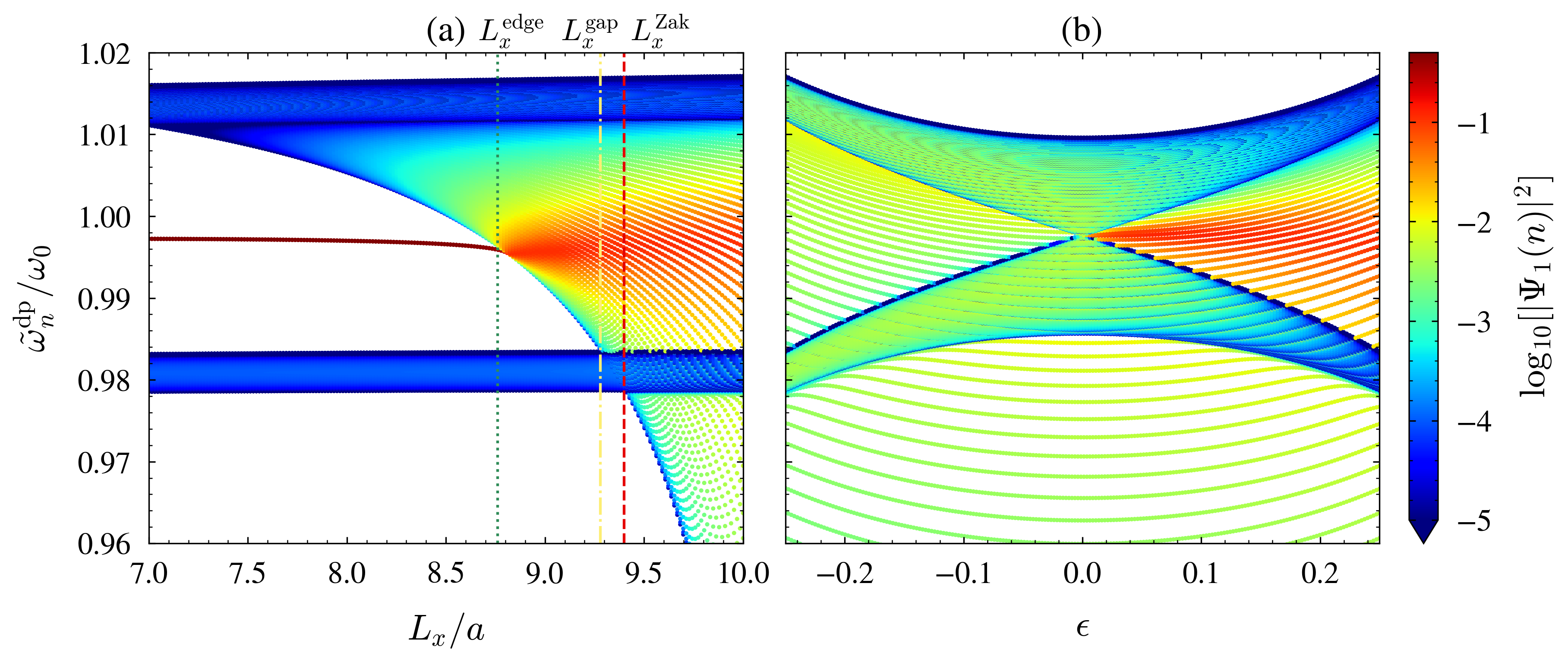}
    \caption{
    Real-space polaritonic eigenfrequencies $\tilde\omega^\mathrm{dp}_n$ (in units of the bare dipole frequency $\omega_0$) as a function of (a) the cavity height $L_x$ 
    and (b) the dimerization parameter $\epsilon$ [Eq.~\eqref{eq:epsilon_dimerization}].
    The color code associated with each eigenstate $n$ represents its probability density on the first dipole site $i=1$, so that it highlights the presence (red) or absence (green or blue) of edge states.
    We fix the dimerization parameter to $\epsilon=0.25$ in panel (a), while the cavity height $L_x=10a$ in panel (b), and we consider a finite chain of $\mathcal{N}=250$ dimers, i.e., $500$ dipoles.
    }
\label{fig:Finite chain - Freq vs Epsilon and Lx}
\end{figure*}

Analogous behaviors have been observed theoretically in similar bipartite systems, from driven ultracold fermions \cite{DiLiberto2014} to zigzag waveguide lattices \cite{Longhi2018}, toy models with next-nearest-neighbor hopping \cite{PerezGonzalez2019}, plasmonic nanoparticles in vacuum \cite{Pocock2019}, or quantum antiferromagnets \cite{Malki2019}.
Using zigzag waveguide lattices, the presence of a nontrivial quantized Zak phase associated with the absence of edge states (that is, what we observe in our model in the lower right region in Fig.~\ref{fig:Zak phase diagram}, with $\epsilon<0$ and $L_x>L_x^\mathrm{Zak}$) has been recently experimentally detected \cite{Jiao2021}.   
The common feature of the systems studied in Refs.~\cite{DiLiberto2014,Longhi2018,PerezGonzalez2019,Pocock2019,Malki2019,Jiao2021} is that there is a coupling parameter breaking the chiral symmetry, which, once enhanced, leads one of the bulk bands to increasingly fill the energy gap.
In our polaritonic system, such parameter is the transverse dimension of the cavity, which allows one to tune the effective photon-mediated dipole-dipole coupling.

%%%%%%%%%%%%%%%%%%%%%%%%%%%%%%%%%%%%%%%%%%%%%%%%%%%%%%%%%%%%
\section{Multiple polaritonic edge states}
\label{sec:Edge states}
%%%%%%%%%%%%%%%%%%%%%%%%%%%%%%%%%%%%%%%%%%%%%%%%%%%%%%%%%%%%

We now move to a discussion of the properties of the finite polaritonic SSH chain.  Importantly, we recall that we here consider a finite chain of dipoles embedded in an infinitely-long cavity, so that we get rid of the effects of the cavity walls in the $z$ direction, and model a \emph{waveguide} cavity instead of a closed one, in contrast to what was done in Ref.~\cite{Downing2019}.
This has a significant impact on the edge states which we study in this section, as briefly discussed in Appendix~\ref{sec:Appendix A: Strong coupling Hamiltonian}.

To determine the spectral properties of the finite system, we write the real-space effective Hamiltonian \eqref{eq:H_dp SW real} in a $2\mathcal{N} \times 2\mathcal{N}$ matrix form using the basis vector $\mathbf{\varphi}^\dagger = (a_1^\dagger,\dots,a_\mathcal{N}^\dagger,b_1^\dagger,\dots,b_\mathcal{N}^\dagger)$, and we numerically diagonalize it to obtain its polaritonic eigenfrequencies $\tilde\omega^\mathrm{dp}_n$ and eigenvectors $\Psi(n) = (\Psi_1(n),\dots,\Psi_{2\mathcal{N}}(n))$, where $n$ labels the eigenvalues in ascending order.

%========================================================
\subsection{Eigenspectrum}

We show the result of the procedure discussed above in Fig.~\ref{fig:Finite chain - Freq vs Epsilon and Lx}, where the eigenfrequencies are plotted as a function of the cavity height $L_x$ in panel (a) and as a function of the dimerization parameter $\epsilon$ in panel (b).
To highlight the presence or absence of edge states, the color code associated with each eigenstate $n$ represents its probability density $|\Psi_1(n)|^2$ on the first site of the chain, on a logarithmic scale.

In Fig.~\ref{fig:Finite chain - Freq vs Epsilon and Lx}(a), we consider a dimerization $\epsilon=0.25$, 
corresponding to the topological sector of the original SSH model.
On the left of the figure, the weak light-matter coupling regime is considered, where the two dipolar bands are only slightly renormalized by the cavity photons [cf. Fig.~\ref{fig:Dispersions SW}(a) for the Fourier-space equivalent]. 
The blue color associated with these two dipolar bands reveals no particular localization on the first site of the chain.
However, two (nearly degenerate) in-gap edge states are visible in dark red, showing their pronounced localization on the first dipole site.
By comparison with the preceding discussion on the bulk topological invariant in Sec.~\ref{sec:Topological phases}, such a situation corresponds to the upper left region of Fig.~\ref{fig:Zak phase diagram}, where a nontrivial Zak phase is found.
We have checked (not shown) that for $\epsilon<0$, there are, as expected, no such edge states. 
In the weak-coupling regime, the bulk-edge correspondence is thus fulfilled, two edge states being present while the two dipolar bands each present a Zak phase of $\tilde{\vartheta}^\mathrm{Zak}=\pi$. 
Nevertheless, we emphasize that these two dipolar edge states are not symmetry-protected topological edge states as the ones found in the original, chiral-symmetric, SSH model.
Indeed, we recall that due to the quasistatic dipole-dipole coupling from Eq.~\eqref{eq:f_nm}, the system does not fulfill chiral symmetry, even in the absence of light-matter coupling.

By increasing the cavity height $L_x$ in Fig.~\ref{fig:Finite chain - Freq vs Epsilon and Lx}(a), the lower, dark band is not affected by the light-matter coupling and its edge localization remains constant. However, the upper band continuously fills the gap with the polaritons that comprise it, the latter arising from the hybridization of the dipoles with the cavity photons, as seen through the avoided crossing scheme with the photonic band in the bulk spectrum [cf.\ Figs.~\ref{fig:Dispersions SW}(b) and \ref{fig:Dispersions SW}(c)].
The two dipolar edge states, however, are only slightly shifted in energy when increasing the cavity height, mainly due to the renormalization of the bare frequency $\omega_0$ into $\tilde{\omega}_0\simeq0.998\omega_0$, [see Eqs.~\eqref{eq: SW renormalized omega0} and \eqref{eq:effective coupling integral I(0)} for an analytical expression of the redshift].
Physically, we attribute this weak change to the fact that the edge states are mainly dark.

As can be seen from Fig.~\ref{fig:Finite chain - Freq vs Epsilon and Lx}(a), the fact that the dipolar edge states and the upper bright band are not similarly shifted in energy as we increase the cavity height allows the polaritons that comprise the latter band to reach the edge-state eigenfrequencies.
Such merging of the dipolar edge states into the bright band arises at a cavity height $L_x^\mathrm{edge}$, marked as a green dotted line.
From this particular cavity height on, we observe the formal disappearance of the two dipolar edge states.
Nevertheless, all of the polaritons belonging to the bright band with an eigenfrequency close to that of the edge states inherit their edge localization, as visible through the red spot on the right of Fig.~\ref{fig:Finite chain - Freq vs Epsilon and Lx}(a), which grows as the cavity height is further increased.
We coin these particular states ``polaritonic edge states."
As we will see in the following, such peculiar states share some of their properties with the original edge states, but also with photonic states originating from the cavity.
From two very localized and nearly degenerate in-gap edge states in the weak-coupling regime, we thus get in the strong-coupling regime numerous polaritonic edge states that are present in a broad frequency range in the bulk of the spectrum.
We insist, however, on the fact that this transition from two dipolar to multiple polaritonic edge states does not represent a TPT.
Indeed, it is not associated with a change of bulk topological invariant, as visible in Fig.~\ref{fig:Zak phase diagram}.

The above-discussed results contrast with what we observe while computing the bulk topological invariant in Sec.~\ref{sec:Topological phases}.
Indeed, for cavity heights $L_x^\mathrm{edge}<L_x<L_x^\mathrm{Zak}$, a nontrivial Zak phase of $\pi$ is found (see the upper left region in Fig.~\ref{fig:Zak phase diagram}), while numerous polaritonic edge states are present. Thus, the bulk-edge correspondence in terms of \emph{number} of edge states is not anymore satisfied.
Moreover, the TPT visible as a red dashed line for $L_x=L_x^\mathrm{Zak}$ does not interfere with the polaritonic edge states which we observe in the finite spectrum of Fig.~\ref{fig:Finite chain - Freq vs Epsilon and Lx}(a).
We attribute the latter breakdown of the bulk-edge correspondence to the fact that the TPT takes place in a system which is already metallic, as $L_x^\mathrm{Zak} > L_x^\mathrm{gap}$, the gap having been closed by the complete chiral symmetry breaking induced by the light-matter coupling.

In Fig.~\ref{fig:Finite chain - Freq vs Epsilon and Lx}(b), we investigate the effect of the dimerization parameter $\epsilon$, and consider the strong light-matter coupling regime with a cavity height $L_x=10a>L_x^\mathrm{Zak}$.
We observe here the absence of edge states when $\epsilon<0$, as it is the case in the usual SSH model.
Looking at the two right frames of the Zak phase diagram in Fig.~\ref{fig:Zak phase diagram}, the bulk invariant indicates, however, a topological phase for $\epsilon<0$.
Therefore, the bulk-edge correspondence is again not verified here.
When $\epsilon>0$, we find ourselves in the case studied previously on the right side of Fig.~\ref{fig:Finite chain - Freq vs Epsilon and Lx}(a), and we observe that increasing the dimerization parameter $\epsilon$ enlarges the energy window in which polaritonic states inherit edge localization.

%========================================================
\subsection{Participation ratio and scaling with the system size}

To study in detail the polaritonic edge states and to maximize the frequency range where they appear, we consider in the following a dimerization $\epsilon=0.25$ and a cavity height $L_x=10a$.
To characterize their localization properties, we use the participation ratio (PR), defined as
\begin{equation}
    \mathrm{PR}(n) = \frac{ \left(  \sum_{i=1}^{2\mathcal{N}}|\Psi_i(n)|^2  \right)^2  }{ \sum_{i=1}^{2\mathcal{N}}|\Psi_i(n)|^4   }.
\label{eq:Participation Ratio}
\end{equation}
Such a quantity provides information on the typical number of dipole sites $i$ occupied by an eigenstate $n$.

\begin{figure}[tb]
    \includegraphics[width=\columnwidth]{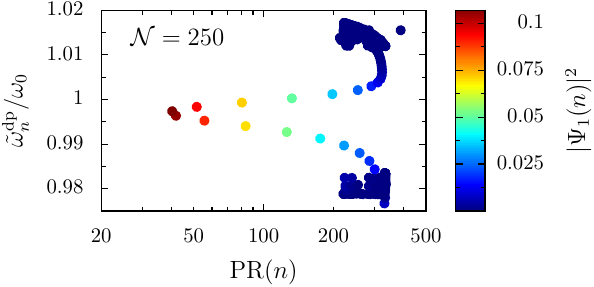}
    \caption{Real-space polaritonic eigenfrequencies $\tilde\omega^\mathrm{dp}_n / \omega_0$ as a function of the participation ratio $\mathrm{PR}(n)$, as defined in Eq.~\eqref{eq:Participation Ratio}.
    The colorscale represents the corresponding probability density on the first site, $i=1$, of the chain.
    In the figure, the cavity height $L_x=10a$, the dimerization parameter $\epsilon=0.25$, and the chain is comprised of $\mathcal{N}=250$ dimers.
    }
\label{fig:spectrum vs PR + Ps1}
\end{figure}

Our results are displayed in Fig.~\ref{fig:spectrum vs PR + Ps1} for a chain of $\mathcal{N}=250$ dimers, where the eigenfrequencies are plotted as a function of the PR on a logarithmic scale, with the color code representing again the probability density at the first site.
Interestingly, the PR of the polaritonic edge states, visible as colored dots from light blue to dark red, follows a bell-shaped curve approximately centered around the eigenfrequency that corresponds to the edge states in the weak-coupling regime ($\tilde{\omega}^\mathrm{dp}_n \simeq 0.998\omega_0$).
Within the parameters used in the figure, we observe here $8$ polaritonic edge states for which at least $\unit[5]{\%}$ of the probability density is found on the first site of the chain only, distributed in a frequency window of about $0.01\omega_0$.
However, their PR contrasts with that of an edge state in the original SSH model or in the weak-coupling regime, here taking large values in between  about $40$ and $200$, instead of approximately $2$.

\begin{figure}[tb]
    \includegraphics[width=\columnwidth]{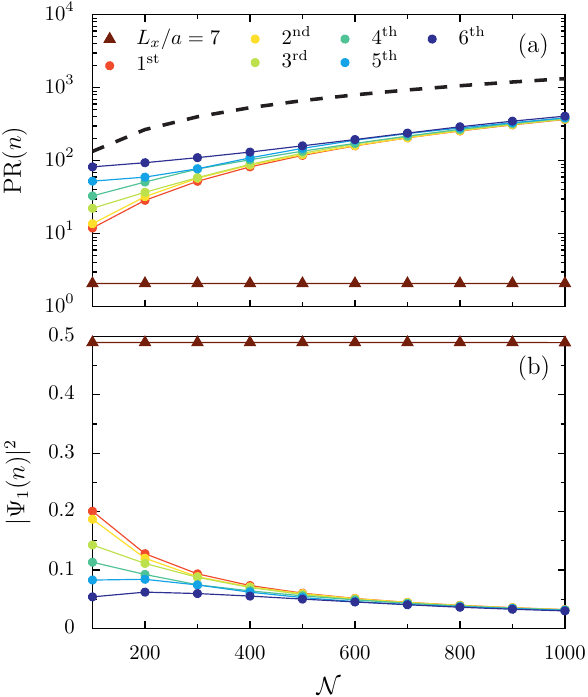}
    \caption{
    Scaling of (a) the participation ratio and (b) the probability density on the first site with the number of dimers $\mathcal{N}$.
    The dark red triangles correspond to the dipolar edge states present in the weak-coupling regime, with a cavity height $L_x=7a$ [cf.\ Fig.~\ref{fig:Finite chain - Freq vs Epsilon and Lx}(a)]. 
    The colored dots correspond to the first six states with the lowest participation ratio in the strong-coupling regime, with a cavity height $L_x=10a$.
    The black dashed line in panel (a) shows the maximum growth rate of the PR for a bipartite chain, $4(\mathcal{N}+1)/3$.
    In the figure, the dimerization parameter $\epsilon=0.25$.
    }
\label{fig:Finite chain PR and Psi1 vs N}
\end{figure}

\begin{figure*}[tb]
    \includegraphics[width=\linewidth]{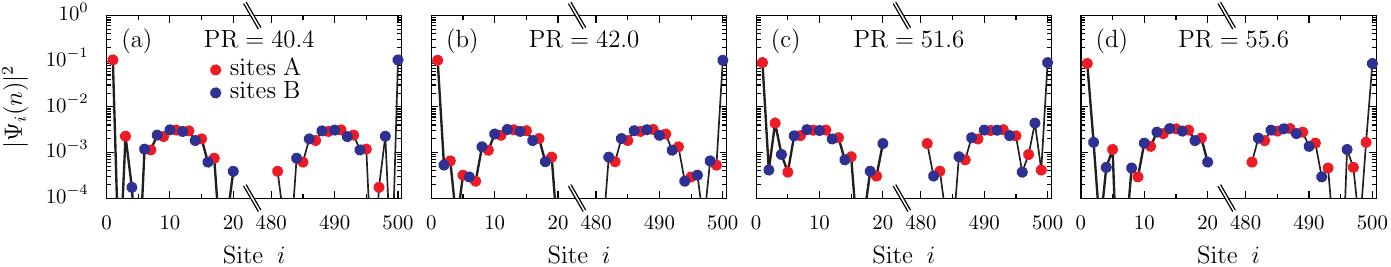}
    \caption{Probability density $|\Psi_i(n)|^2$ along the sites $i$ of a chain comprised of $\mathcal{N}=250$ dimers, for the states $n$ with (a) the first, (b) second, (c) third, and (d) fourth lowest participation ratio.
    The red and blue dots represent, respectively, sites belonging to the $A$ and $B$ sublattices.
    The cavity height $L_x=10a$ and the dimerization $\epsilon=0.25$.}
\label{fig:Proba densities}
\end{figure*}

This difference is illustrated in Fig.~\ref{fig:Finite chain PR and Psi1 vs N}(a), where we plot the scaling of the participation ratio $\mathrm{PR}(n)$ with the number of dimers $\mathcal{N}$, for the six polaritonic edge states with the lowest PR (colored dots), as well as for the two dipolar edge states present in the weak-coupling regime [visible on the left of Fig.~\ref{fig:Finite chain - Freq vs Epsilon and Lx}(a)], for a cavity height $L_x=7a$ (dark red triangles).
Extended states, i.e., states with a localization length larger than the system, are characterized by a PR scaling linearly with the number of dimers $\mathcal{N}$, while the PR of states that are formally localized must be size-independent.
Following such classification, one observes in Fig.~\ref{fig:Finite chain PR and Psi1 vs N}(a) that the polaritonic edge states are not formally localized, their PR scaling with the system size, with a growth rate approaching the maximal one for a bipartite chain, $4(\mathcal{N}+1)/3$, shown as a black dashed line.
While for small chain sizes a clear difference in PR is visible between each polaritonic edge state, such a dissimilarity fades out when the number of dimers $\mathcal{N}$ increases.
We explain this behavior by the increasing number of polaritonic edge states when the size of the chain increases.
Indeed, as the density of states with an eigenfrequency around $\tilde{\omega}_0$ increases, the number of polaritonic states which resonate with the original edge states, and hence which inherit their edge localization, grows.
The six states we show in Fig.~\ref{fig:Finite chain PR and Psi1 vs N}(a) are therefore more and more similar as $\mathcal{N}$ increases.
Such scaling with the PR is drastically different from the one we observe in the weak-coupling regime (dark red triangles), in which the edge states have a constant PR of about $2$, as it is the case in the original SSH model \cite{Asboth2016}.
This key difference between edge states in the weak- and strong-coupling regimes originates solely from the fact that in the latter case, polaritonic edge states feature a significant bulk part, induced by their hybridization with the cavity photons, and are therefore no longer only localized on the ends of the dipole chain, so that their localization length increases naturally with the size of the chain.

In Fig.~\ref{fig:Finite chain PR and Psi1 vs N}(b), we present our results for the scaling of the probability density on the first site of the chain $|\Psi_1(n)|^2$ with the number of dimers $\mathcal{N}$, for the 
same states as in Fig.~\ref{fig:Finite chain PR and Psi1 vs N}(a).
In the weak-coupling regime, as in the original SSH model, almost half of the probability density of the edge states is located on the first site of the chain (the other half being localized on the last site), independently of the system size.
In the strong-coupling regime, however, the situation is less usual.
While for small system sizes, a fifth of the probability density of some polaritonic edge states can be located on the first site of the chain only, such fraction decreases and tends towards a constant value, around a few percents, when the chain becomes longer.
Similarly to the scaling of the PR, such a behavior is explained by the growing number of polaritonic edge states when the system size increases. 
This leads the edge localization to be shared between more and more polaritonic edge states.
Interestingly, we will demonstrate in Sec.~\ref{sec:Transport} that despite this diffusion of the edge localization between numerous states, driven-dissipative transport simulations with lossy dipoles, taking into account the linewidth of the excitations, allow these polaritonic edge states to be probed.

%========================================================
\subsection{Polaritonic edge states}

To conclude this section, and before studying the transport properties of the polaritonic edge states, we here discuss their shape in real space.
To this end, the probability density along a chain of $\mathcal{N}=250$ dimers is shown in Fig.~\ref{fig:Proba densities} in the strong-coupling regime ($L_x=10a$) and for $\epsilon=0.25$, for the four most localized states with the lowest PR.
The probability density is represented on a logarithmic scale, while the horizontal axis has been truncated to focus on the edges of the chain.
The red (blue) points correspond to sites belonging to the $A$ ($B$) sublattice. 
One observes that these four polaritonic states display a clear localization on their edges, with more than an order of magnitude higher than their probability density in the bulk of the chain.
Two distinct regimes of spatial extensions are apparent on each of these states.
First, on the first few and last few sites, the states are exponentially localized on the edges, with for some of the states an alternation between sites $A$ and $B$, as is the case of the chirally-symmetric topological edge states of the original SSH model.
Second, in the bulk of the chain we observe a non-negligible probability density, evenly distributed along the chain, similarly to extended plane waves.
While the first regime is reminiscent of topological edge states of the original SSH model, the second demonstrates the polaritonic nature of these states, having inherited the delocalized, plane-wave nature of the cavity photons.

%%%%%%%%%%%%%%%%%%%%%%%%%%%%%%%%%%%%%%%%%%%%%%%%%%%%%%%%%%%%
\section{Steady-state transport}
\label{sec:Transport}
%%%%%%%%%%%%%%%%%%%%%%%%%%%%%%%%%%%%%%%%%%%%%%%%%%%%%%%%%%%%

To relate the results obtained in Sec.~\ref{sec:Edge states} with measurable quantities, here we investigate the transport of excitations in the polaritonic SSH model.
Specifically, we focus on the potential contribution of edge states on energy transport throughout the chain, which has been recently unveiled experimentally \cite{Chang2022_arXiv} and theoretically investigated with the help of a simplified Tavis-Cummings model \cite{Wei2022}.

We consider a driven-dissipative scenario by adding the driving term
\begin{equation}
    H_\mathrm{drive}(t) = \hbar\Omega_\mathrm{R}\sin(\omega_\mathrm{d}t)\left( a^{\phantom{\dagger}}_1 + a^{\dagger}_1 \right)
    \label{eq: Hdrive}
\end{equation}
to the effective Hamiltonian \eqref{eq:H_dp SW real}.
The equation above models the continuous illumination of the first dipole site by a transversely polarized monochromatic electric field with amplitude $E_0$ and driving frequency $\omega_\mathrm{d}$, with $\Omega_\mathrm{R} = E_0\sqrt{Q^2/2M\hbar\omega_0}$ the corresponding Rabi frequency.
We assume that the dynamics can be described by the Lindblad master equation for the density matrix 
\begin{align}
    \dot{\rho} =&\; \frac{\mathrm{i}}{\hbar}\left[ \rho, \tilde{H}_\mathrm{dp}+H_{\mathrm{drive}}(t) \right] \nonumber \\
    &- \frac{\gamma}{2}\sum_{m=1}^{\mathcal{N}} \left(  \left\{ a_m^\dagger a_m^{\phantom{\dagger}} + b_m^\dagger b_m^{\phantom{\dagger}}, \rho \right\}   - 2a_m^{\phantom{\dagger}} \rho a_m^\dagger - 2b_m^{\phantom{\dagger}} \rho b_m^\dagger \right).
    \label{eq:Lindblad master equation}
\end{align}
Here, the damping rate $\gamma$ quantifies the influence of 
a phenomenological Markovian bath responsible for the dissipation of the dipolar emitters. Dissipation typically originates from radiative and Ohmic losses, and in this section, we consider $\gamma=0.002\omega_0$.
Such a narrow linewidth can be achieved experimentally using emitters with weak losses, such as, e.g., microwave resonators or dielectric and SiC nanoparticles \cite{Mann2018,Wang2018b}. 

To characterize the excitation of an emitter belonging to the $A$ ($B$) sublattice, we introduce its dimensionless dipole moment $p_m^A = \langle a_m^{\phantom\dagger} + a_m^\dagger \rangle$ ($p_m^B = \langle b_m^{\phantom\dagger} + b_m^\dagger \rangle$).
Solving the master equation \eqref{eq:Lindblad master equation}, we obtain the steady-state amplitudes $|p_i|$ bared by a dipole on the site $i$ of the chain, which belongs either to the $A$ or $B$ sublattice.
We note that these dimensionless amplitudes are proportional to the square root of the power radiated in the far field by a dipole, through the classical Larmor formula \cite{Jackson2007,Allard2022}.

\begin{figure}[b]
    \includegraphics[width=\linewidth]{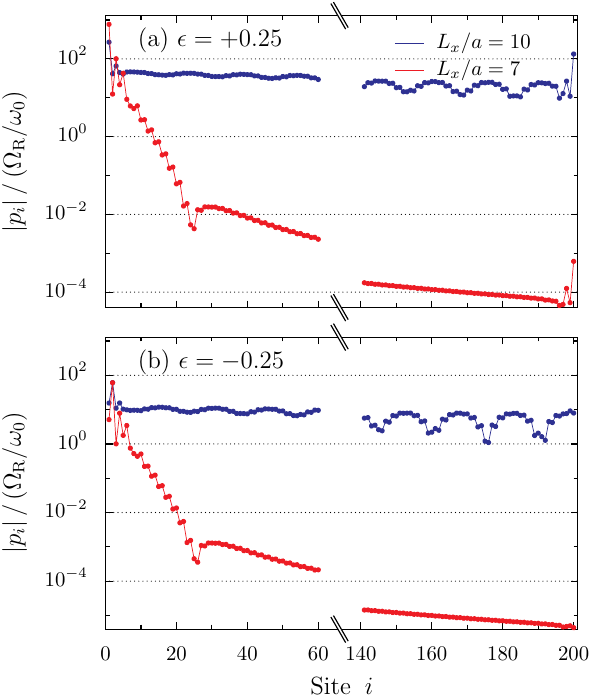}
    \caption{Steady-state amplitude of the dipole moment $|p_i|$ (in units of $\Omega_\mathrm{R}/\omega_0$) as a function of the dipole site $i$, for a chain with dimerization (a) $\epsilon=+0.25$ and (b) $\epsilon=-0.25$.
    The red (blue) symbols correspond to the weak- (strong-)coupling regime, with a cavity height $L_x=7a$ ($L_x=10a$).
    The propagation results from a monochromatic drive on the first dipole site at a frequency $\omega_\mathrm{d}=0.998\omega_0$. The chain is comprised of $\mathcal{N}=100$ dimers with damping rates $\gamma=0.002\omega_0$.}
\label{fig:Transport along chain}
\end{figure}

Figure~\ref{fig:Transport along chain} displays our findings for the steady-state amplitudes $|p_i|$ of the dipole moments, scaled by the Rabi frequency in units of the bare dipole frequency, $\Omega_\mathrm{R}/\omega_0$, and as a function of the sites $i$ of a chain comprised of $\mathcal{N}=100$ dimers.
The first site is driven at a frequency $\omega_\mathrm{d} = 0.998\omega_0 \simeq \tilde{\omega}_0$, corresponding approximately to the edge state eigenfrequencies in the weak-coupling regime.
The propagation signals are shown in a log-linear plot for both the weak- ($L_x/a=7$) and strong-coupling ($L_x/a=10$)  regimes, by red and blue symbols, respectively. 

In Fig.~\ref{fig:Transport along chain}(a) we consider a dimerization parameter $\epsilon=+0.25$.
Edge states are clearly visible in both coupling regimes, through a large rise of the excitation at the end of the chain, the dipole moment increasing there by one order of magnitude.
In the first few sites of the chain, the propagation signal quickly decays for both coupling regimes, following an exponential decay which is reminiscent of the nearest-neighbor dipole-dipole coupling \cite{BrandstetterKunc2016}.
However, for longer distances, the transport characteristics are very distinct.
On the one hand, for $L_x/a=7$ (red dots), we observe a steep quasi-exponential decay, induced by the light-matter coupling, followed by an algebraic tail decaying with the inverse distance cubed, arising from the quasistatic dipole-dipole coupling.
On the other hand, for $L_x/a=10$ (blue dots), the propagation follows an exponential decay with a large decay length, rendering the decay profile nearly flat.
This second exponential decay originates solely from the cavity-induced effective dipole-dipole coupling (see Sec.~\ref{sec:Model C} and Appendix \ref{sec:Appendix C: Effective coupling integrals}).
Such a decay is physically explained by the hybridization of the bright, upper dipolar band with cavity photons, and stands for the polaritonic cavity-enhanced transport.
The slope in a log-linear plot of such a cavity-induced exponential decay is dictated by both the damping rate $\gamma$ and the cavity height $L_x$, becoming flatter as the latter increases. 
Therefore, the driving of the polaritonic edge states plotted as blue dots in Fig.~\ref{fig:Transport along chain}(a) presents interesting transport characteristics, allowing for efficient end-to-end edge state transport, as opposed to what is observed in red for the dipolar edge state.
We note that for very long chains or more lossy dipoles, the algebraic tail is also present in the polaritonic transport.

In Fig.~\ref{fig:Transport along chain}(b), we study the propagation along the chain when the first dipole is driven at the same frequency $\omega_\mathrm{d}=0.998\omega_0$, now for $\epsilon=-0.25$.
As discussed in the previous section, such negative value of the dimerization $\epsilon$ is associated with the absence of edge states.
The same transport regimes as in Fig.~\ref{fig:Transport along chain}(a) are observed, but no rise of the dipole moment is found at the end of the chain.
Moreover, there is an overall decrease of the dipole moment amplitudes along the chain as compared to the $\epsilon=+0.25$ case.
This is in agreement with the so-called dimerization-assisted transport that has been studied in detail in Ref.~\cite{Wei2022}.

\begin{figure}[tb]
    \includegraphics[width=\linewidth]{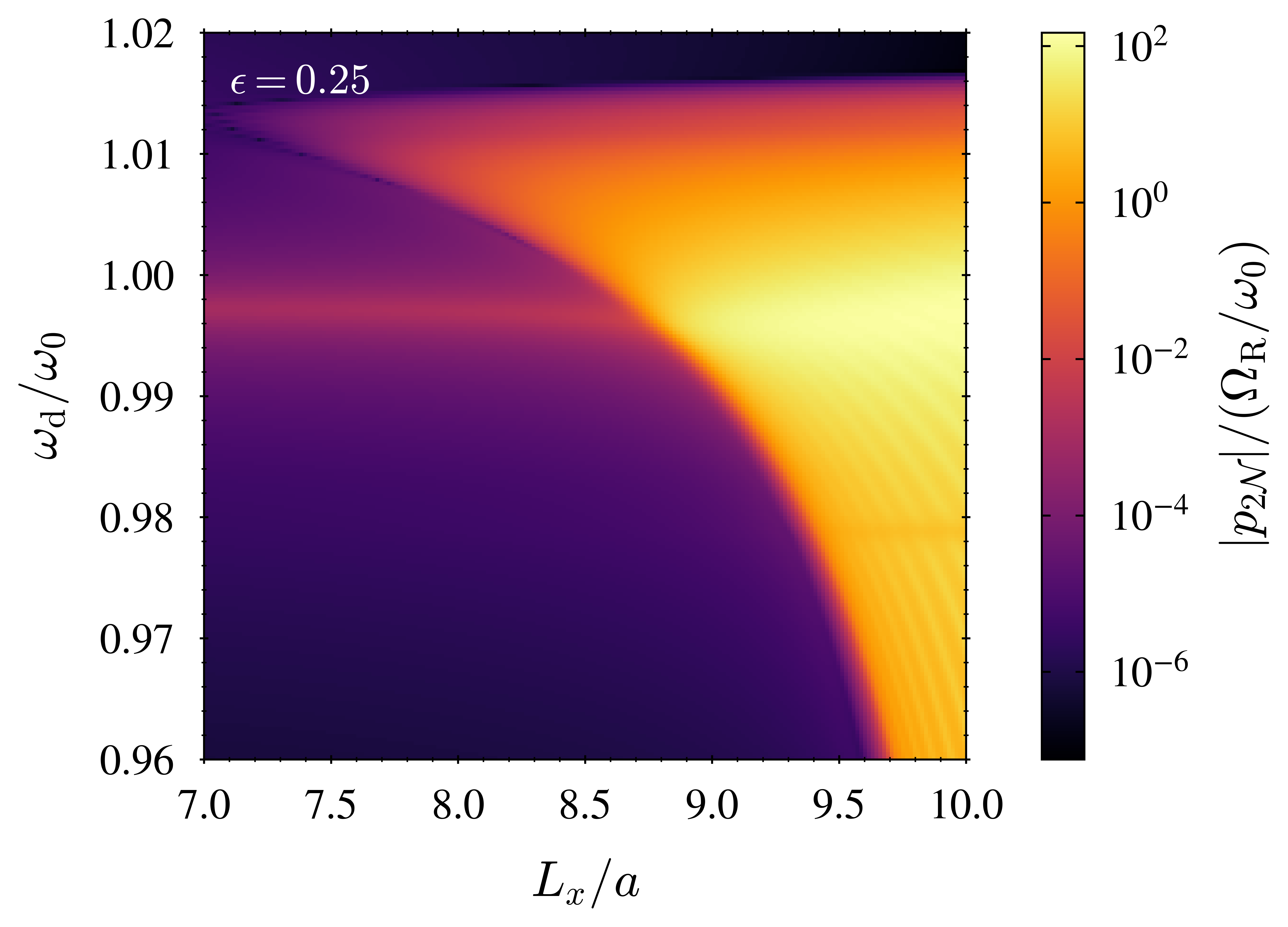}
    \caption{Steady-state amplitude of the last dipole moment of the chain $|p_{2\mathcal{N}}|$ (in units of $\Omega_\mathrm{R}/\omega_0$), for increasing cavity heights $L_x/a$ and driving frequencies $\omega_\mathrm{d}/\omega_0$.
    In the figure, $\mathcal{N}=100$, $\epsilon=0.25$, and $\gamma=0.002\omega_0$.}
\label{fig:Transport spectrum heatmap}
\end{figure}

To highlight the effect of the cavity on end-to-end transport, we display in Fig.~\ref{fig:Transport spectrum heatmap} a density plot of the normalized steady-state amplitude of the last dipole moment of the chain, $|p_{2\mathcal{N}}|/(\Omega_\mathrm{R}/\omega_0)$, using a logarithmic scale, as a function of both the cavity height $L_x$ and the driving frequency $\omega_\mathrm{d}$.
A chain of $\mathcal{N}=100$ dimers with a dimerization $\epsilon=0.25$ is considered.
Interestingly, Fig.~\ref{fig:Transport spectrum heatmap} has similarities with Fig.~\ref{fig:Finite chain - Freq vs Epsilon and Lx}(a), which shows the probability density of the eigenstates at one end of the chain as a function of the cavity height, with a dimerization parameter also fixed to the same value as here, so that we recover traces of the spectrum in our transport simulations.
The light orange layer on the right of Fig.~\ref{fig:Transport spectrum heatmap} in the strong-coupling regime corresponds to the driving of the polaritons that originate from the bright, upper dipolar band which is continuously redshifted when the cavity height is increased [as seen in Fig.~\ref{fig:Finite chain - Freq vs Epsilon and Lx}(a)].
Owing from their polaritonic nature, they feature enhanced transport characteristics, notably through the cavity-induced exponential decay discussed above in Fig.~\ref{fig:Transport along chain}, explaining their large dipole moment amplitude at the end of the chain.
In contrast, when driving at a frequency corresponding to the bare dipolar bands in the weak-coupling regime [around $\omega_\mathrm{d}/\omega_0=0.980$ and $\omega_\mathrm{d}/\omega_0=1.015$, cf. Fig.~\ref{fig:Finite chain - Freq vs Epsilon and Lx}(a)], only a small dipole moment is found on the last site of the chain, demonstrating poor transport properties.

Examining driving frequencies around $\omega_\mathrm{d}/\omega_0 = 1$ in Fig.~\ref{fig:Transport spectrum heatmap}, we observe the presence of the edge states, which show a particularly large dipole moment at the end of the chain.
Notably, in the weak-coupling regime, here for cavity heights $L_x \lesssim L_x^\mathrm{edge} \simeq 8.7a$, the dipolar edge states appear as a dark orange beam in the center left of Fig.~\ref{fig:Transport spectrum heatmap}.
In the strong-coupling regime, here for cavity heights $L_x \gtrsim L_x^\mathrm{edge}$, the polaritonic edge states are visible through the bright yellow spot in the center right of Fig.~\ref{fig:Transport spectrum heatmap}.
Crucially, we observe that the latter bright spot spreads over a broad range of driving frequencies.
Therefore, in addition to allowing very efficient edge state transport between the two ends of the chain, the cavity also largely broadens the edge state frequency band.

%%%%%%%%%%%%%%%%%%%%%%%%%%%%%%%%%%%%%%%%%%%%%%%%%%%%%%%%%%%%
\section{Robustness to disorder}
\label{sec:Disorder}
%%%%%%%%%%%%%%%%%%%%%%%%%%%%%%%%%%%%%%%%%%%%%%%%%%%%%%%%%%%%

To complement our study of the cavity-induced polaritonic edge states featured by the polaritonic SSH model \eqref{eq:Hamiltonian}, we discuss here their robustness to disorder.
Specifically, we study the effect of disorder in the intra- and interdimer spacings $d_1$ and $d_2$ (see Fig.~\ref{fig: sketch}), which corresponds to off-diagonal disorder. We consider these spacings to be uncorrelated random variables uniformly distributed  within the interval $[d_{1,2}(1-\Delta) , d_{1,2}(1+\Delta)]$, where the dimensionless parameter $\Delta$ is the amplitude of the spacing fluctuations and characterizes the disorder strength.

Interestingly, the off-diagonal disorder which we consider does not break the chiral symmetry of the bipartite chain, so that it does not alter the topological edge states of the original (chirally symmetric) SSH model \cite{Asboth2016}.
However, as discussed in Sec.~\ref{sec:Model C}, due to the dipole-dipole couplings beyond nearest neighbors the polaritonic SSH model does not fulfill chiral symmetry, in both the weak- and strong-coupling regimes.
On the one hand, although being reminiscent of the chiral symmetry of the original SSH model, the polaritonic edge states should therefore not present any formal robustness against off-diagonal disorder.
On the other hand, polaritons, through their photonic part, have been proven robust against disorder, presenting a cavity-protection effect \cite{Houdre1996,Michetti_2005}.
The interplay between topological polaritonic edge states and disorder is thereby highly nontrivial.

\begin{figure}[tb]
    \includegraphics[width=\linewidth]{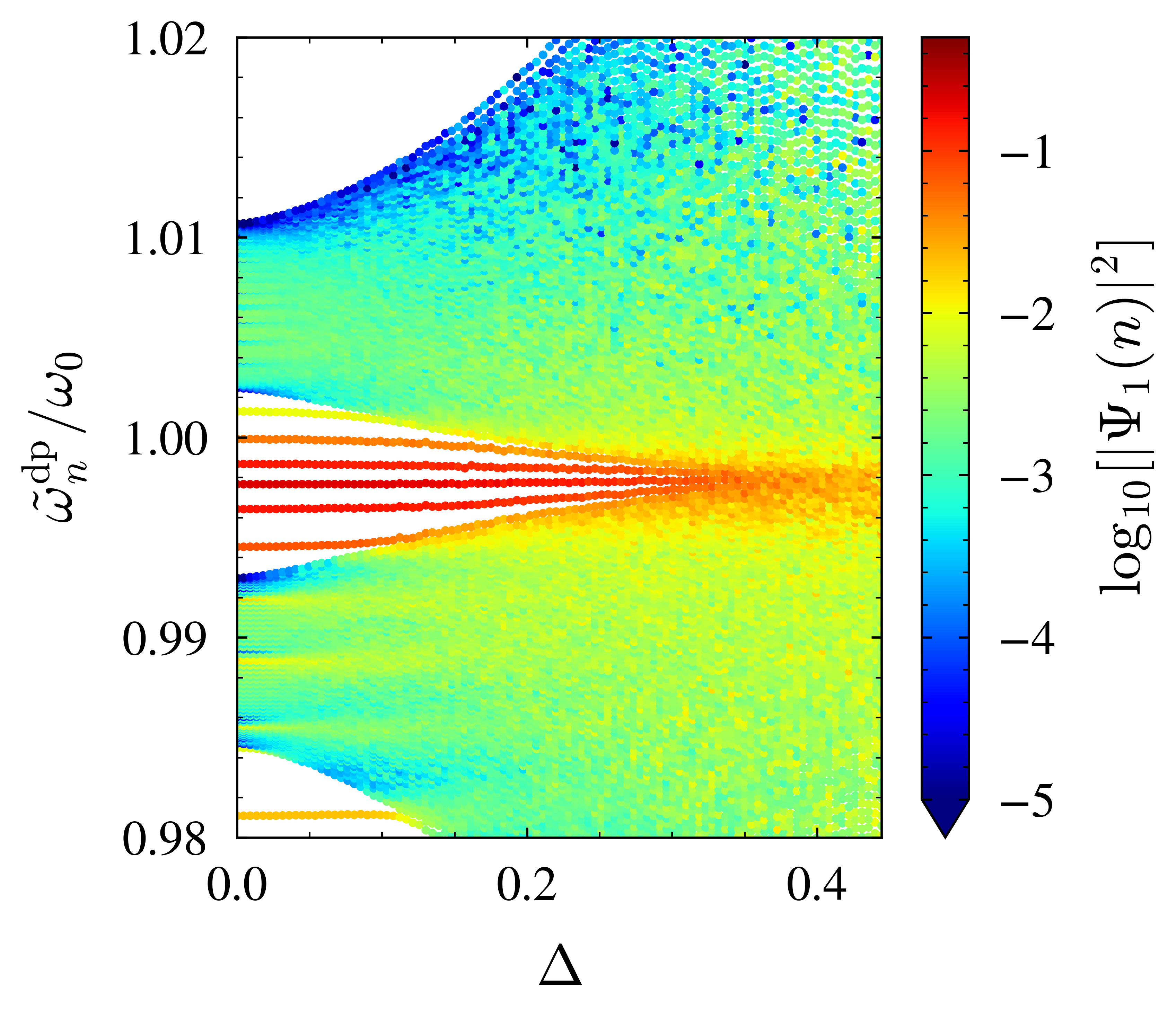}
    \caption{Real-space polaritonic eigenfrequencies $\tilde\omega^\mathrm{dp}_n$ (in units of the bare dipole frequency $\omega_0$) as a function of the dimensionless disorder strength $\Delta$.
    The color code associated with each eigenstate $n$ represents its probability density on the first dipole site $i=1$, so that it highlights the presence (red) or absence (green or blue) of edge states.
    A chain of $\mathcal{N}=100$ dimers with an average value of the dimerization parameter $\epsilon=0.1$ embedded in a waveguide cavity with height $L_x=10a$ is considered, and the data have been averaged over $100$ disorder realizations.}
\label{fig:off-diagonal disorder}
\end{figure}

We clarify this point by plotting in Fig.~\ref{fig:off-diagonal disorder} the disorder-averaged real-space polaritonic eigenfrequencies $\tilde\omega^\mathrm{dp}_n/\omega_0$ as a function of the dimensionless disorder strength $\Delta$, for a chain of $\mathcal{N}=100$ dimers with an average dimerization parameter $\epsilon=0.1$, embedded in a cavity with height $L_x=10a$.
As in Fig.~\ref{fig:Finite chain - Freq vs Epsilon and Lx}, the color code displays, on a logarithmic scale, the probability density on the first dipole site $i=1$ associated with each eigenstate $n$, so that it reveals the presence (red) or absence (green or blue) of states that are highly localized at the two ends of the chain.

We observe in Fig.~\ref{fig:off-diagonal disorder} the characteristic bandwidth widening as the disorder strength increases.
Importantly, only the dipolar eigenstates, corresponding to the eigenfrequencies of the bare dipolar bands, here around $\tilde\omega^\mathrm{dp}_n/\omega_0=0.990$ and $\tilde\omega^\mathrm{dp}_n/\omega_0=1.005$, undergo this effect.
In contrast, the eigenfrequencies of the polaritonic states, that are highly renormalized by the cavity photons, remain constant on average at small disorder strength.
The latter polaritons, showing the cavity-protection effect, are visible in Fig.~\ref{fig:off-diagonal disorder} through the yellow stripes in the lower left region of the figure, as well as through the red stripes, showing the polaritonic edge states, around $\tilde\omega^\mathrm{dp}_n/\omega_0=1$.
Such robustness of the polaritons against disorder fades out as they merge into the dipolar bands.
As anticipated by the broken chiral symmetry, the polaritonic edge states (red stripes in Fig.~\ref{fig:off-diagonal disorder}) are therefore not formally robust against off-diagonal disorder, but can survive at high levels of disorder, with a large probability density of $0.1$ on the first site up to $\Delta \simeq 0.25$ with the parameters considered in the figure.
We attribute this tolerance to disorder both to their polaritonic nature and to their topological origin, being reminiscent of the topologically protected edge states of the original, chiral-symmetric SSH model.
Indeed, dipolar edge states present in the weak-coupling regime, although not being polaritonic, also show very good tolerance to off-diagonal disorder, up to the closing of the bandgap by the bandwidth increase led by disorder (not shown).

Interestingly, once the off-diagonal disorder is strong enough to let the polaritonic edge states merge into the dipolar bands, the dipolar states with eigenfrequencies close to that of the edge states inherit part of their edge localization, as apparent through the orange spot in the center right of Fig.~\ref{fig:off-diagonal disorder}.
This mechanism is very similar in nature to what gives rise to the polaritonic edge states, as we described in Sec.~\ref{sec:Edge states}, i.e., the fact that edge and bulk states are not similarly affected by the increase of a given parameter (here the positional disorder strength $\Delta$, in Sec.~\ref{sec:Edge states} the cavity height $L_x$).

To conclude this section, we note that even a small amount of disorder in the frequency of the bare dipoles destroys the edge-state localization \cite{Allard2023_PhD}, in contrast with what we observe here for interdipole spacing disorder.

%%%%%%%%%%%%%%%%%%%%%%%%%%%%%%%%%%%%%%%%%%%%%%%%%%%%%%%%%%%%
\section{Conclusion}
\label{sec:Conclusions}
%%%%%%%%%%%%%%%%%%%%%%%%%%%%%%%%%%%%%%%%%%%%%%%%%%%%%%%%%%%%

In summary, we have analyzed in detail the effect of strong light-matter coupling on topological edge states, studying the eigenspectrum, the eigenstates, as well as the transport properties of a bipartite chain of emitters (modeled as point dipoles) strongly coupled to a multimode waveguide cavity.
Such a system mimics a variation of the celebrated two-band Su-Schrieffer-Heeger model, with the addition of an effective dipole-dipole coupling mediated by the cavity photons.
We have found such cavity-mediated coupling to take the form of an exponential decay whose 
decay length increases as one enters in the strong-coupling regime.

We have shown that the effect of the strong light-matter coupling is to hybridize and redshift the bright dipolar band into a polaritonic one, which strongly breaks the chiral symmetry of the model, and can close the energy gap, so that the system becomes metallic.
In this regime, a cavity-induced topological phase transition, i.e., a change in the bulk topological invariant of the system, is observed.
We find that such a transition, which takes place in a gapless regime, is not associated with the appearance nor disappearance of edge states, leading the bulk-edge correspondence not to be fulfilled.

In the topological sector of the original SSH model, the strong-coupling regime leads the in-gap edge states to merge into the polaritonic bulk band.
We have unveiled that even if the formal in-gap edge states are thus destroyed, all the polaritons entering in resonance with the edge states inherit part of their localization properties.
Edge localization is then diffused into multiple polaritonic edge states that keep a delocalized bulk part and cover a wide frequency range.
Our results highlight the peculiar properties of these polaritonic edge states, in particular, the latters taking advantage of their polaritonic nature to allow efficient energy transport between the two ends of the chain.
Moreover, the broadening of the edge-state frequency band makes them sensitive to a wide range of driving frequencies.
These two unusual cavity-induced effects on topological edge states may facilitate their experimental detection.

Furthermore, by studying the impact of disorder on the interdipole spacings, we have demonstrated the tolerance of the polaritonic edge states to disorder.
Thanks both to their polaritonic nature and topological origin, polaritonic edge states being reminiscent of the symmetry-protected edge states of the original SSH model, they can survive at high levels of off-diagonal disorder.

Our model, allowing a detailed numerical and partly analytical understanding of the strong light-matter coupling effects on topological edge states, could constitute a building block of a more general theory of topological polaritonics, essential to the successful implementation of topological photonic technologies.
A direct and attractive extension of our present model is its generalization to ultra- and deep-strong light-matter couplings, the latter allowing surprising quantum effects \cite{Kockum2019_review,FornDiaz2019,Masuki_arxiv_2022}.

%%%%%%%%%%%%%%%%%%%%%%%%%%%%%%%%%%%%%%%%%%%%%%%%%%%%%%%%%%%%
%\section*{Acknowledgments}
%%%%%%%%%%%%%%%%%%%%%%%%%%%%%%%%%%%%%%%%%%%%%%%%%%%%%%%%%%%%
\begin{acknowledgments}
We are greatful to Cl\'ement Tauber for valuable and stimulating discussions. We thank Paloma Arroyo Huidobro, Denis Basko, and Jean-No\"el Fuchs for insightful comments, as well as Charles A.\ Downing for his careful reading of our manuscript. This work of the Interdisciplinary Thematic Institute QMat, as part of the ITI 2021-2028 program of the University of Strasbourg, CNRS, and Inserm, was supported by IdEx Unistra (ANR 10 IDEX 0002), and by SFRI STRAT’US Projects No.\ ANR-20-SFRI-0012 and No.\ ANR-17-EURE-0024 under the framework of the French Investments for the Future Program.
\end{acknowledgments}

%----------------------------------------------------------%
%----------------------------------------------------------%
%----------------------------------------------------------%
%----------------------------------------------------------%
%----------------------------------------------------------%

\appendix
%%%%%%%%%%%%%%%%%%%%%%%%%%%%%%%%%%%%%%%%%%%%%%%%%%%%%%%%%%%%%%%%%%%%%%%%%%%%%%%%%%%%
\section{Strong-coupling Hamiltonian}
\label{sec:Appendix A: Strong coupling Hamiltonian}
%%%%%%%%%%%%%%%%%%%%%%%%%%%%%%%%%%%%%%%%%%%%%%%%%%%%%%%%%%%%%%%%%%%%%%%%%%%%%%%%%%%%

In the main text, we model the light-matter coupling between the dipolar chain and the multimode optical cavity by the Hamiltonian \eqref{eq:H_dp-ph pbc}.
In this appendix, we briefly give details on its derivation.

In the Coulomb gauge, emitters couple to photonic degrees of freedom via the usual minimal light-matter coupling Hamiltonian (in cgs units)
\begin{equation}
    H_{\mathrm{dp}\textrm{-}\mathrm{ph}} = \frac{Q}{Mc} \sum_{s=A,B}\sum_{m=1}^{\mathcal{N}} \mathbf{\Pi}_m^s \cdot \mathbf{A} (\mathbf{r}_m^s),
\label{eq:H_dp-ph general}
\end{equation}
which, importantly, mediates the retarded part of the Coulomb interaction between the dipoles \cite{Craig2012}.
Equation~\eqref{eq:H_dp-ph general} couples the momenta 
\begin{subequations}
\label{eq:Momentum}
\begin{align}
    \mathbf{\Pi}_m^A&=\mathrm{i}\sqrt{\frac{M\hbar\omega_0}{2}}\left({a_m}^\dagger - a_m^{\phantom{\dagger}}\right)\hat{x},\\
    \mathbf{\Pi}_m^B&=\mathrm{i}\sqrt{\frac{M\hbar\omega_0}{2}}\left({b_m}^\dagger - b_m^{\phantom{\dagger}}\right)\hat{x},
\end{align}
\end{subequations}
associated with the $m$th dipole excitation belonging to the $A$ or $B$ sublattice to the vector potential 
\begin{equation}
    \mathbf{A} (\mathbf{r}_m^s) = \sum_{\mathbf{k}, l, \hat\lambda_{\mathbf{k},l}} \sqrt{ \frac{2\pi\hbar c^2}{\omega^{\mathrm{ph}}_{\mathbf{k},l}} }\left[ \mathbf{f}_{\mathbf{k},l}^{\hat{\lambda}_{\mathbf{k},l}}(\mathbf{r}_m^s)c_{\mathbf{k},l}^{\hat{\lambda}_{\mathbf{k},l}} + \mathrm{H.c.} \right]%+ {\mathbf{f}_{\mathbf{k},l}^{\hat{\lambda}_{\mathbf{k},l}*}}(\mathbf{r}_m^s){c_{\mathbf{k},l}^{\hat{\lambda}_{\mathbf{k},l}\dagger}}  \right]
    \label{eq:gauge potential}
\end{equation}
of the quantized electromagnetic field, 
evaluated at the position of each dipole $\mathbf{r}_m^s = (L_x/2 , L_y/2, z_m^s)$, which we place at the center of the cavity.
Here, the $z$ coordinate of a dipole $m$ belonging to the $A$ or $B$ sublattice reads $z_m^A = (m-1)d + d_\mathrm{cav}$ or $z_m^B = (m-1)d + d_\mathrm{cav} + d_1$, respectively, with $d_\mathrm{cav}$ the distance of the first and last dipole to the ends of the cavity in the $z$ direction.
In Eq.~\eqref{eq:gauge potential}, the bosonic ladder operators $c_{\mathbf{k},l}^{\hat{\lambda}_{\mathbf{k},l}}$ and ${c_{\mathbf{k},l}^{\hat{\lambda}_{\mathbf{k},l}\dagger}}$ respectively annihilate and create a cavity photon with wavevector $\mathbf{k}$, Umklapp band index $l$, and transverse polarization $\hat{\lambda}_{\mathbf{k},l}$
, while the mode functions $\mathbf{f}_{\mathbf{k},l}^{\hat{\lambda}_{\mathbf{k},l}}(\mathbf{r}_m^s)$ depend on the cavity geometry and boundary conditions.
As mentioned in Sec.~\ref{sec:Model B} and discussed in further detail in Refs.~\cite{Downing2019,Downing2021}, choosing the geometry $L_y = 3L_x$ and $3a \lesssim L_x \lesssim 20a$ allows us to consider only the lowest photonic band and a single photon polarization, for which we therefore drop the associated index.

Aside from the above-discussed approximation, we note that the diamagnetic, so-called ``$A^2$'' term is neglected in the Hamiltonian \eqref{eq:H_dp-ph general}, since its impact on the spectrum is only of qualitative importance in the case of the ultra-strong or deep-strong coupling regimes \cite{Kockum2019_review, Stefano2019}, in which we do not enter in this study.

While Ref.~\cite{Downing2019}, as well as other recent studies \cite{Downing2021,Allard2022}, considered hard wall boundaries for the cavity in the three space directions, here we opt for periodic boundary conditions in the $z$ direction.
Such a choice of boundary conditions, with the fact that only the band $(n_x,n_y,q)=(0,1,q)$ is taken into account, leads to the mode function \cite{Kakazu1994} 
\begin{equation}
    \mathbf{f}_{q,l}(z_m^s) = \sqrt{\frac{2}{L_x L_y L_z}} \, \mathrm{e}^{\mathrm{i} q_l z_m^s} \, {\hat x}
    \label{eq:mode function after approx}
\end{equation}
with $q_l=q-2\pi l/d$,
which, once combined with Eqs.~\eqref{eq:H_dp-ph general}--\eqref{eq:gauge potential}, gives within the rotating-wave approximation the coupling Hamiltonian \eqref{eq:H_dp-ph pbc} presented in the main text. 

We note that the longitudinal photonic wavenumber is given by $q = 2 \pi p/L_z$ with $p \in [-L_z/2d , +L_z/2d]$, so that $q$ belongs to the first Brillouin zone, and $L_z = L_\mathrm{chain} + 2d_\mathrm{cav}$, where the length of the chain $L_\mathrm{chain}=(\mathcal{N}-1)d+d_1$.
In order for our periodic boundary conditions to be physically relevant, $d_\mathrm{cav}$ must go to infinity, as done in Sec.~\ref{sec:Model C} by considering the continuous limit for $q$.
In that sense, our model here is equivalent to a \emph{waveguide} cavity, with open ends far away enough from the first and last dipoles.

Interestingly, such a distance between the cavity ends and the chain qualitatively affects the localization of the eigenstates of the system at the two ends of the chain.
Indeed, the finite spectrum of the polaritonic SSH Hamiltonian in a waveguide cavity \eqref{eq:H_dp SW real} and, in particular, the polaritonic edge states which we study throughout this paper, considerably differ from what was found in Ref.~\cite{Downing2019}, where a closed cuboidal cavity with $d_\mathrm{cav}=d-d_{1}/2$ was considered.
With walls that close from the ends of the chain, edge states can localize not only on the dipoles, but also on the cavity walls, allowing them to be more robust to the strong light-matter coupling, and to diffuse much less in the bulk.
We verified that as long as $d_\mathrm{cav} \gtrsim 10d$,  such boundary effects become insignificant \cite{Allard2023_PhD}.
We found that similar non-negligible boundary effects are also at the origin of the Tamm edge states that have been unveiled in Ref.~\cite{Downing2021} in a regular chain embedded in a closed cuboidal cavity using a similar model as the one employed here.

%%%%%%%%%%%%%%%%%%%%%%%%%%%%%%%%%%%%%%%%%%%%%%%%%%%%%%%%%%%%%%%%%%%%%%%%%%%%%%%%%%%%
\section{Schrieffer-Wolff transformation}
\label{sec:Appendix B: Schrieffer-Wolff transformation}
%%%%%%%%%%%%%%%%%%%%%%%%%%%%%%%%%%%%%%%%%%%%%%%%%%%%%%%%%%%%%%%%%%%%%%%%%%%%%%%%%%%%

In reciprocal space, the Bloch Hamiltonian stemming from Eq.~\eqref{eq:Hamiltonian} can be represented by a $3\times3$ matrix, containing both dipolar and photonic degrees of freedom. 
Focusing on the fate of the two dipolar bands coupled to the cavity modes, 
we integrate out the photonic degrees of freedom of the full polaritonic Hamiltonian \eqref{eq:Hamiltonian} by 
performing the Schrieffer-Wolff unitary transformation \cite{Schrieffer1966}
\begin{equation}
    \tilde{H} = \mathrm{e}^S H \mathrm{e}^{-S} \simeq H + [S , H] + \frac{1}{2}[ S , [ S , H ] ] 
    \label{eq:SW transformation}
\end{equation}
Here, using the fact that the quasistatic dipole-dipole coupling strength $\Omega/\omega_0 \ll 1$, the anti-Hermitian operator $S$ (i.e., $S^\dagger=-S$) is determined such that
\begin{equation}
    [S , H_\mathrm{dp}\left(\Omega=0\right) + H_\mathrm{ph}] = - H_{\mathrm{dp}\textrm{-}\mathrm{ph}},
\label{eq:SW condition}
\end{equation}
so as to eliminate coupling terms of the order of $\Omega\xi_{q,l}^2/{\omega_0}^3$ in the effective Hamiltonian. From the condition \eqref{eq:SW condition}, we find 
\begin{align}
    S =& -\mathrm{i} \sum_{m=1}^\mathcal{N} \sum_{q,l} \sqrt{\frac{d}{L_z}} \frac{\xi_{q,l}}{\omega^\mathrm{ph}_{q,l} - \omega_0} \nonumber \\
    &\times \left[  \mathrm{e}^{\mathrm{i}mq_ld} c_{q,l}^{\phantom{\dagger}} \left( a_m^\dagger\mathrm{e}^{-\mathrm{i}\chi_{q,l}} + b_m^\dagger\mathrm{e}^{\mathrm{i}\chi_{q,l}}  \right) + \mathrm{H.c.}  \right].
\end{align}
The dipolar and photonic subspaces are then decoupled to second order in the light-matter coupling strength \eqref{eq:light-matter coupling strength pbc},
\begin{equation}
    \tilde{H} \simeq {H}_\mathrm{dp} + {H}_\mathrm{ph} + \frac{1}{2}[S,H_{\mathrm{dp}\textrm{-}\mathrm{ph}}] \equiv \tilde{H}_\mathrm{dp} + \tilde{H}_\mathrm{ph}.
    \label{eq:SW effective Hamiltonian}
\end{equation}
Computing the commutator in Eq.~\eqref{eq:SW effective Hamiltonian} and focusing on the dipolar subspace, we obtain the effective Hamiltonian \eqref{eq:H_dp SW real}.

%%%%%%%%%%%%%%%%%%%%%%%%%%%%%%%%%%%%%%%%%%%%%%%%%%%%%%%%%%%%%%%%%%%%%%%%%%%%%%%%%%%%
\section{Effective coupling integrals}
\label{sec:Appendix C: Effective coupling integrals}
%%%%%%%%%%%%%%%%%%%%%%%%%%%%%%%%%%%%%%%%%%%%%%%%%%%%%%%%%%%%%%%%%%%%%%%%%%%%%%%%%%%%

The cavity-induced renormalizations of the bare resonance frequency and of the intra- and intersublattice dipole-dipole coupling, derived in Sec.~\ref{sec:Model C} [see Eq.~\eqref{eq: SW renormalized quantities real}], appear as
\begin{equation}
  \frac{d}{2\pi}\sum_{l=-\infty}^{+\infty}\int_{-\pi/d}^{+\pi/d} \mathrm{d}q \frac{\xi^2_{q,l}}{\omega_{q,l}^\mathrm{ph} - \omega_0}\,\mathrm{e}^{\mathrm{i}\alpha q_ld} = \frac{\omega_0 a^3}{L_x L_y} \mathcal{I}(\alpha),
   \label{eq:effective coupling integral}
\end{equation}
with integrals of the form
\begin{align}
    \mathcal{I}(\alpha) &= \sum_{l=-\infty}^{+\infty} \int_{-\pi/d}^{+\pi/d} \mathrm{d}q\, \frac{\omega_0^2\,  \mathrm{e}^{\mathrm{i}\alpha q_ld}}{ \omega^\mathrm{ph}_{q,l} ( \omega^\mathrm{ph}_{q,l}   -   \omega_0 ) } \nonumber \\
    &= \int_{-\infty}^{+\infty} \mathrm{d}q \, \frac{\omega_0^2\, \mathrm{e}^{\mathrm{i}\alpha qd}}{ \omega^\mathrm{ph}_{q,0} ( \omega^\mathrm{ph}_{q,0}   -   \omega_0 ) } ,
    \label{eq:effective coupling integral I(alpha) def}
\end{align}
where $\alpha=0$, $m-m'$, or $m-m'-d_1/d$.

The particular case of $\mathcal{I}(0)$ can be readily evaluated.
It appears in the cavity-renormalized onsite frequency [cf. Eq.~\eqref{eq: SW renormalized omega0}], which therefore reads
\begin{align}
    \tilde{\omega}_0 =&\; \omega_0 - \frac{2\omega_0 a^2 k_0a }{L_x L_y} \frac{\omega_0}{\sqrt{(\omega^\mathrm{ph}_{0,0})^2 - {\omega_0}^2}} \nonumber \\
    &\times \left[ \arctan\left( \frac{\omega_0}{\sqrt{(\omega^\mathrm{ph}_{0,0})^2 - {\omega_0}^2}} \right) + \frac{\pi}{2} \right].
    \label{eq:effective coupling integral I(0)}
\end{align}
Such a term only leads to a slight shift of the bare frequency $\omega_0$ (less than about $\unit[0.3]{\%}$), which increases with the cavity height $L_x$ when the photon frequency $\omega^\mathrm{ph}_{0,0}$ approaches the bare frequency $\omega_0$.

For the intra- and intersublattice coupling renormalizations, where $\alpha$ equals, respectively, $m-m'$ and $m-m'-d_1/d$, we use the fact that in our perturbation theory the photonic frequency remains larger than the bare dipole one, so that we can rewrite the effective coupling integrals $\mathcal{I}(\alpha)$ as
\begin{align}
    \mathcal{I}(\alpha) = \sum_{p=2}^\infty\int_{-\infty}^{+\infty} \mathrm{d}q \left( \frac{\omega_0}{\omega^\mathrm{ph}_{q,0}} \right)^p \mathrm{e}^{\mathrm{i}\alpha qd} \equiv \sum_{p=2}^\infty \mathcal{J}_p(\alpha),
    \label{eq:effective coupling integral J(alpha)}
\end{align}
which can be evaluated as an infinite sum of modified Bessel functions of the second kind $K_{\nu}(z)$, 
\begin{equation}
    \mathcal{J}_p(\alpha) = \frac{\sqrt{\pi}(k_0d)^p}{\Gamma(p/2)d}\left( \frac{|\alpha|L_y}{2\pi d} \right)^{\frac{p-1}{2}} K_{\frac{p-1}{2}} \left( |\alpha|\frac{\pi d}{L_y} \right), 
    \label{eq:effective coupling integral J_p(alpha)}
\end{equation}
where $\Gamma(z)$ represents the gamma function.
In all the results presented in this work involving the finite spectrum, namely, Figs.~\ref{fig:Finite chain - Freq vs Epsilon and Lx}--\ref{fig:off-diagonal disorder}, we truncated such infinite sum to $p_\mathrm{max}=100$, having checked the irrelevance of higher-order terms.
Each term in the sum of Eq.~\eqref{eq:effective coupling integral J_p(alpha)} corresponds to a nearly exponential decay and the sum is dominated by the first, $p=2$ term
\begin{equation}
    \mathcal{J}_2(\alpha) = \frac{(k_0d)^2 L_y}{2d^2}\exp{\left(-|\alpha|\frac{\pi d}{L_y}\right)},
    \label{eq:effective coupling integral J_2}
\end{equation}
a pure exponential decay with a clear dependence on the cavity width $L_y=3L_x$.
Increasing the cavity height, i.e., entering in the strong-coupling regime, leads the latter exponential decay to fall on larger distances, so that the cavity-induced effective dipole-dipole coupling becomes stronger.
Such an (almost) exponential decay induced by the strong-coupling regime is directly visible through the steady-state transport along the dipole chain, as studied in Sec.~\ref{sec:Transport}.
It allows for a large effective dipole-dipole coupling at intermediate distances.
At very long distances, however, such cavity-induced exponential decay is superseded by the quasistatic dipolar couplings \eqref{eq:lattice sums real} which decay as one over the distance cubed, as discussed in Sec.~\ref{sec:Transport}.

\bibliography{refs}

%apsrev4-2.bst 2019-01-14 (MD) hand-edited version of apsrev4-1.bst
%Control: key (0)
%Control: author (8) initials jnrlst
%Control: editor formatted (1) identically to author
%Control: production of article title (0) allowed
%Control: page (0) single
%Control: year (1) truncated
%Control: production of eprint (0) enabled
\begin{thebibliography}{69}%
\makeatletter
\providecommand \@ifxundefined [1]{%
 \@ifx{#1\undefined}
}%
\providecommand \@ifnum [1]{%
 \ifnum #1\expandafter \@firstoftwo
 \else \expandafter \@secondoftwo
 \fi
}%
\providecommand \@ifx [1]{%
 \ifx #1\expandafter \@firstoftwo
 \else \expandafter \@secondoftwo
 \fi
}%
\providecommand \natexlab [1]{#1}%
\providecommand \enquote  [1]{``#1''}%
\providecommand \bibnamefont  [1]{#1}%
\providecommand \bibfnamefont [1]{#1}%
\providecommand \citenamefont [1]{#1}%
\providecommand \href@noop [0]{\@secondoftwo}%
\providecommand \href [0]{\begingroup \@sanitize@url \@href}%
\providecommand \@href[1]{\@@startlink{#1}\@@href}%
\providecommand \@@href[1]{\endgroup#1\@@endlink}%
\providecommand \@sanitize@url [0]{\catcode `\\12\catcode `\$12\catcode `\&12\catcode `\#12\catcode `\^12\catcode `\_12\catcode `\%12\relax}%
\providecommand \@@startlink[1]{}%
\providecommand \@@endlink[0]{}%
\providecommand \url  [0]{\begingroup\@sanitize@url \@url }%
\providecommand \@url [1]{\endgroup\@href {#1}{\urlprefix }}%
\providecommand \urlprefix  [0]{URL }%
\providecommand \Eprint [0]{\href }%
\providecommand \doibase [0]{https://doi.org/}%
\providecommand \selectlanguage [0]{\@gobble}%
\providecommand \bibinfo  [0]{\@secondoftwo}%
\providecommand \bibfield  [0]{\@secondoftwo}%
\providecommand \translation [1]{[#1]}%
\providecommand \BibitemOpen [0]{}%
\providecommand \bibitemStop [0]{}%
\providecommand \bibitemNoStop [0]{.\EOS\space}%
\providecommand \EOS [0]{\spacefactor3000\relax}%
\providecommand \BibitemShut  [1]{\csname bibitem#1\endcsname}%
\let\auto@bib@innerbib\@empty
%</preamble>
\bibitem [{\citenamefont {Ozawa}\ \emph {et~al.}(2019)\citenamefont {Ozawa}, \citenamefont {Price}, \citenamefont {Amo}, \citenamefont {Goldman}, \citenamefont {Hafezi}, \citenamefont {Lu}, \citenamefont {Rechtsman}, \citenamefont {Schuster}, \citenamefont {Simon}, \citenamefont {Zilberberg},\ and\ \citenamefont {Carusotto}}]{Ozawa2019}%
  \BibitemOpen
  \bibfield  {author} {\bibinfo {author} {\bibfnamefont {T.}~\bibnamefont {Ozawa}}, \bibinfo {author} {\bibfnamefont {H.~M.}\ \bibnamefont {Price}}, \bibinfo {author} {\bibfnamefont {A.}~\bibnamefont {Amo}}, \bibinfo {author} {\bibfnamefont {N.}~\bibnamefont {Goldman}}, \bibinfo {author} {\bibfnamefont {M.}~\bibnamefont {Hafezi}}, \bibinfo {author} {\bibfnamefont {L.}~\bibnamefont {Lu}}, \bibinfo {author} {\bibfnamefont {M.~C.}\ \bibnamefont {Rechtsman}}, \bibinfo {author} {\bibfnamefont {D.}~\bibnamefont {Schuster}}, \bibinfo {author} {\bibfnamefont {J.}~\bibnamefont {Simon}}, \bibinfo {author} {\bibfnamefont {O.}~\bibnamefont {Zilberberg}},\ and\ \bibinfo {author} {\bibfnamefont {I.}~\bibnamefont {Carusotto}},\ }\bibfield  {title} {\bibinfo {title} {Topological photonics},\ }\href {https://doi.org/10.1103/RevModPhys.91.015006} {\bibfield  {journal} {\bibinfo  {journal} {Rev. Mod. Phys.}\ }\textbf {\bibinfo {volume} {91}},\ \bibinfo {pages} {015006} (\bibinfo {year} {2019})}\BibitemShut {NoStop}%
\bibitem [{\citenamefont {Rider}\ \emph {et~al.}(2019)\citenamefont {Rider}, \citenamefont {Palmer}, \citenamefont {Pocock}, \citenamefont {Xiao}, \citenamefont {Huidobro},\ and\ \citenamefont {Giannini}}]{Rider2019}%
  \BibitemOpen
  \bibfield  {author} {\bibinfo {author} {\bibfnamefont {M.~S.}\ \bibnamefont {Rider}}, \bibinfo {author} {\bibfnamefont {S.~J.}\ \bibnamefont {Palmer}}, \bibinfo {author} {\bibfnamefont {S.~R.}\ \bibnamefont {Pocock}}, \bibinfo {author} {\bibfnamefont {X.}~\bibnamefont {Xiao}}, \bibinfo {author} {\bibfnamefont {P.~A.}\ \bibnamefont {Huidobro}},\ and\ \bibinfo {author} {\bibfnamefont {V.}~\bibnamefont {Giannini}},\ }\bibfield  {title} {\bibinfo {title} {A perspective on topological nanophotonics: Current status and future challenges},\ }\href {https://doi.org/10.1063/1.5086433} {\bibfield  {journal} {\bibinfo  {journal} {J. Appl. Phys.}\ }\textbf {\bibinfo {volume} {125}},\ \bibinfo {pages} {120901} (\bibinfo {year} {2019})}\BibitemShut {NoStop}%
\bibitem [{\citenamefont {Rider}\ \emph {et~al.}(2022)\citenamefont {Rider}, \citenamefont {Buend\'ia}, \citenamefont {Abujetas}, \citenamefont {Huidobro}, \citenamefont {Sánchez-Gil},\ and\ \citenamefont {Giannini}}]{Rider2022}%
  \BibitemOpen
  \bibfield  {author} {\bibinfo {author} {\bibfnamefont {M.~S.}\ \bibnamefont {Rider}}, \bibinfo {author} {\bibfnamefont {A.}~\bibnamefont {Buend\'ia}}, \bibinfo {author} {\bibfnamefont {D.~R.}\ \bibnamefont {Abujetas}}, \bibinfo {author} {\bibfnamefont {P.~A.}\ \bibnamefont {Huidobro}}, \bibinfo {author} {\bibfnamefont {J.~A.}\ \bibnamefont {Sánchez-Gil}},\ and\ \bibinfo {author} {\bibfnamefont {V.}~\bibnamefont {Giannini}},\ }\bibfield  {title} {\bibinfo {title} {Advances and prospects in topological nanoparticle photonics},\ }\href {https://doi.org/10.1021/acsphotonics.1c01874} {\bibfield  {journal} {\bibinfo  {journal} {ACS Photonics}\ }\textbf {\bibinfo {volume} {9}},\ \bibinfo {pages} {1483} (\bibinfo {year} {2022})}\BibitemShut {NoStop}%
\bibitem [{\citenamefont {Hasan}\ and\ \citenamefont {Kane}(2010)}]{HasanReview2010}%
  \BibitemOpen
  \bibfield  {author} {\bibinfo {author} {\bibfnamefont {M.~Z.}\ \bibnamefont {Hasan}}\ and\ \bibinfo {author} {\bibfnamefont {C.~L.}\ \bibnamefont {Kane}},\ }\bibfield  {title} {\bibinfo {title} {Colloquium: Topological insulators},\ }\href {https://doi.org/10.1103/RevModPhys.82.3045} {\bibfield  {journal} {\bibinfo  {journal} {Rev. Mod. Phys.}\ }\textbf {\bibinfo {volume} {82}},\ \bibinfo {pages} {3045} (\bibinfo {year} {2010})}\BibitemShut {NoStop}%
\bibitem [{\citenamefont {Qi}\ and\ \citenamefont {Zhang}(2011)}]{Xiao-LiangReview2011}%
  \BibitemOpen
  \bibfield  {author} {\bibinfo {author} {\bibfnamefont {X.-L.}\ \bibnamefont {Qi}}\ and\ \bibinfo {author} {\bibfnamefont {S.-C.}\ \bibnamefont {Zhang}},\ }\bibfield  {title} {\bibinfo {title} {Topological insulators and superconductors},\ }\href {https://doi.org/10.1103/RevModPhys.83.1057} {\bibfield  {journal} {\bibinfo  {journal} {Rev. Mod. Phys.}\ }\textbf {\bibinfo {volume} {83}},\ \bibinfo {pages} {1057} (\bibinfo {year} {2011})}\BibitemShut {NoStop}%
\bibitem [{\citenamefont {Asbóth}\ \emph {et~al.}(2016)\citenamefont {Asbóth}, \citenamefont {Oroszlány},\ and\ \citenamefont {Pályi}}]{Asboth2016}%
  \BibitemOpen
  \bibfield  {author} {\bibinfo {author} {\bibfnamefont {J.~K.}\ \bibnamefont {Asbóth}}, \bibinfo {author} {\bibfnamefont {L.}~\bibnamefont {Oroszlány}},\ and\ \bibinfo {author} {\bibfnamefont {A.}~\bibnamefont {Pályi}},\ }\href {https://doi.org/10.1007/978-3-319-25607-8} {\emph {\bibinfo {title} {A Short Course on Topological Insulators}}}\ (\bibinfo  {publisher} {Springer},\ \bibinfo {address} {Berlin},\ \bibinfo {year} {2016})\BibitemShut {NoStop}%
\bibitem [{\citenamefont {Garcia-Vidal}\ \emph {et~al.}(2021)\citenamefont {Garcia-Vidal}, \citenamefont {Ciuti},\ and\ \citenamefont {Ebbesen}}]{EbbesenReview2021}%
  \BibitemOpen
  \bibfield  {author} {\bibinfo {author} {\bibfnamefont {F.~J.}\ \bibnamefont {Garcia-Vidal}}, \bibinfo {author} {\bibfnamefont {C.}~\bibnamefont {Ciuti}},\ and\ \bibinfo {author} {\bibfnamefont {T.~W.}\ \bibnamefont {Ebbesen}},\ }\bibfield  {title} {\bibinfo {title} {Manipulating matter by strong coupling to vacuum fields},\ }\href {https://doi.org/10.1126/science.abd0336} {\bibfield  {journal} {\bibinfo  {journal} {Science}\ }\textbf {\bibinfo {volume} {373}},\ \bibinfo {pages} {eabd0336} (\bibinfo {year} {2021})}\BibitemShut {NoStop}%
\bibitem [{\citenamefont {Appugliese}\ \emph {et~al.}(2022)\citenamefont {Appugliese}, \citenamefont {Enkner}, \citenamefont {Paravicini-Bagliani}, \citenamefont {Beck}, \citenamefont {Reichl}, \citenamefont {Wegscheider}, \citenamefont {Scalari}, \citenamefont {Ciuti},\ and\ \citenamefont {Faist}}]{Appugliese2022}%
  \BibitemOpen
  \bibfield  {author} {\bibinfo {author} {\bibfnamefont {F.}~\bibnamefont {Appugliese}}, \bibinfo {author} {\bibfnamefont {J.}~\bibnamefont {Enkner}}, \bibinfo {author} {\bibfnamefont {G.~L.}\ \bibnamefont {Paravicini-Bagliani}}, \bibinfo {author} {\bibfnamefont {M.}~\bibnamefont {Beck}}, \bibinfo {author} {\bibfnamefont {C.}~\bibnamefont {Reichl}}, \bibinfo {author} {\bibfnamefont {W.}~\bibnamefont {Wegscheider}}, \bibinfo {author} {\bibfnamefont {G.}~\bibnamefont {Scalari}}, \bibinfo {author} {\bibfnamefont {C.}~\bibnamefont {Ciuti}},\ and\ \bibinfo {author} {\bibfnamefont {J.}~\bibnamefont {Faist}},\ }\bibfield  {title} {\bibinfo {title} {Breakdown of topological protection by cavity vacuum fields in the integer quantum {Hall} effect},\ }\href {https://doi.org/10.1126/science.abl5818} {\bibfield  {journal} {\bibinfo  {journal} {Science}\ }\textbf {\bibinfo {volume} {375}},\ \bibinfo {pages} {1030} (\bibinfo {year} {2022})}\BibitemShut {NoStop}%
\bibitem [{\citenamefont {Su}\ \emph {et~al.}(1979)\citenamefont {Su}, \citenamefont {Schrieffer},\ and\ \citenamefont {Heeger}}]{Su1979}%
  \BibitemOpen
  \bibfield  {author} {\bibinfo {author} {\bibfnamefont {W.~P.}\ \bibnamefont {Su}}, \bibinfo {author} {\bibfnamefont {J.~R.}\ \bibnamefont {Schrieffer}},\ and\ \bibinfo {author} {\bibfnamefont {A.~J.}\ \bibnamefont {Heeger}},\ }\bibfield  {title} {\bibinfo {title} {Solitons in polyacetylene},\ }\href {https://doi.org/10.1103/PhysRevLett.42.1698} {\bibfield  {journal} {\bibinfo  {journal} {Phys. Rev. Lett.}\ }\textbf {\bibinfo {volume} {42}},\ \bibinfo {pages} {1698} (\bibinfo {year} {1979})}\BibitemShut {NoStop}%
\bibitem [{\citenamefont {Di~Liberto}\ \emph {et~al.}(2014)\citenamefont {Di~Liberto}, \citenamefont {Malpetti}, \citenamefont {Japaridze},\ and\ \citenamefont {Morais~Smith}}]{DiLiberto2014}%
  \BibitemOpen
  \bibfield  {author} {\bibinfo {author} {\bibfnamefont {M.}~\bibnamefont {Di~Liberto}}, \bibinfo {author} {\bibfnamefont {D.}~\bibnamefont {Malpetti}}, \bibinfo {author} {\bibfnamefont {G.~I.}\ \bibnamefont {Japaridze}},\ and\ \bibinfo {author} {\bibfnamefont {C.}~\bibnamefont {Morais~Smith}},\ }\bibfield  {title} {\bibinfo {title} {Ultracold fermions in a one-dimensional bipartite optical lattice: Metal-insulator transitions driven by shaking},\ }\href {https://doi.org/10.1103/PhysRevA.90.023634} {\bibfield  {journal} {\bibinfo  {journal} {Phys. Rev. A}\ }\textbf {\bibinfo {volume} {90}},\ \bibinfo {pages} {023634} (\bibinfo {year} {2014})}\BibitemShut {NoStop}%
\bibitem [{\citenamefont {Wang}\ and\ \citenamefont {Zhao}(2018{\natexlab{a}})}]{Wang2018b}%
  \BibitemOpen
  \bibfield  {author} {\bibinfo {author} {\bibfnamefont {B.~X.}\ \bibnamefont {Wang}}\ and\ \bibinfo {author} {\bibfnamefont {C.~Y.}\ \bibnamefont {Zhao}},\ }\bibfield  {title} {\bibinfo {title} {Topological phonon polaritons in one-dimensional {non-Hermitian} silicon carbide nanoparticle chains},\ }\href {https://doi.org/10.1103/PhysRevB.98.165435} {\bibfield  {journal} {\bibinfo  {journal} {Phys. Rev. B}\ }\textbf {\bibinfo {volume} {98}},\ \bibinfo {pages} {165435} (\bibinfo {year} {2018}{\natexlab{a}})}\BibitemShut {NoStop}%
\bibitem [{\citenamefont {Longhi}(2018)}]{Longhi2018}%
  \BibitemOpen
  \bibfield  {author} {\bibinfo {author} {\bibfnamefont {S.}~\bibnamefont {Longhi}},\ }\bibfield  {title} {\bibinfo {title} {Probing one-dimensional topological phases in waveguide lattices with broken chiral symmetry},\ }\href {https://doi.org/10.1364/OL.43.004639} {\bibfield  {journal} {\bibinfo  {journal} {Opt. Lett.}\ }\textbf {\bibinfo {volume} {43}},\ \bibinfo {pages} {4639} (\bibinfo {year} {2018})}\BibitemShut {NoStop}%
\bibitem [{\citenamefont {Downing}\ and\ \citenamefont {Weick}(2018)}]{Downing2018}%
  \BibitemOpen
  \bibfield  {author} {\bibinfo {author} {\bibfnamefont {C.~A.}\ \bibnamefont {Downing}}\ and\ \bibinfo {author} {\bibfnamefont {G.}~\bibnamefont {Weick}},\ }\bibfield  {title} {\bibinfo {title} {Topological plasmons in dimerized chains of nanoparticles: {Robustness} against long-range quasistatic interactions and retardation effects},\ }\href {https://doi.org/10.1140/epjb/e2018-90199-0} {\bibfield  {journal} {\bibinfo  {journal} {Eur. Phys. J. B}\ }\textbf {\bibinfo {volume} {91}},\ \bibinfo {pages} {253} (\bibinfo {year} {2018})}\BibitemShut {NoStop}%
\bibitem [{\citenamefont {P\'erez-Gonz\'alez}\ \emph {et~al.}(2019)\citenamefont {P\'erez-Gonz\'alez}, \citenamefont {Bello}, \citenamefont {G\'omez-Le\'on},\ and\ \citenamefont {Platero}}]{PerezGonzalez2019}%
  \BibitemOpen
  \bibfield  {author} {\bibinfo {author} {\bibfnamefont {B.}~\bibnamefont {P\'erez-Gonz\'alez}}, \bibinfo {author} {\bibfnamefont {M.}~\bibnamefont {Bello}}, \bibinfo {author} {\bibfnamefont {A.}~\bibnamefont {G\'omez-Le\'on}},\ and\ \bibinfo {author} {\bibfnamefont {G.}~\bibnamefont {Platero}},\ }\bibfield  {title} {\bibinfo {title} {Interplay between long-range hopping and disorder in topological systems},\ }\href {https://doi.org/10.1103/PhysRevB.99.035146} {\bibfield  {journal} {\bibinfo  {journal} {Phys. Rev. B}\ }\textbf {\bibinfo {volume} {99}},\ \bibinfo {pages} {035146} (\bibinfo {year} {2019})}\BibitemShut {NoStop}%
\bibitem [{\citenamefont {Pocock}\ \emph {et~al.}(2019)\citenamefont {Pocock}, \citenamefont {Huidobro},\ and\ \citenamefont {Giannini}}]{Pocock2019}%
  \BibitemOpen
  \bibfield  {author} {\bibinfo {author} {\bibfnamefont {S.~R.}\ \bibnamefont {Pocock}}, \bibinfo {author} {\bibfnamefont {P.~A.}\ \bibnamefont {Huidobro}},\ and\ \bibinfo {author} {\bibfnamefont {V.}~\bibnamefont {Giannini}},\ }\bibfield  {title} {\bibinfo {title} {Bulk-edge correspondence and long-range hopping in the topological plasmonic chain},\ }\href {https://doi.org/10.1515/nanoph-2019-0033} {\bibfield  {journal} {\bibinfo  {journal} {Nanophotonics}\ }\textbf {\bibinfo {volume} {8}},\ \bibinfo {pages} {1337} (\bibinfo {year} {2019})}\BibitemShut {NoStop}%
\bibitem [{\citenamefont {Bello}\ \emph {et~al.}(2019)\citenamefont {Bello}, \citenamefont {Platero}, \citenamefont {Cirac},\ and\ \citenamefont {González-Tudela}}]{Bello2019}%
  \BibitemOpen
  \bibfield  {author} {\bibinfo {author} {\bibfnamefont {M.}~\bibnamefont {Bello}}, \bibinfo {author} {\bibfnamefont {G.}~\bibnamefont {Platero}}, \bibinfo {author} {\bibfnamefont {J.~I.}\ \bibnamefont {Cirac}},\ and\ \bibinfo {author} {\bibfnamefont {A.}~\bibnamefont {González-Tudela}},\ }\bibfield  {title} {\bibinfo {title} {Unconventional quantum optics in topological waveguide {QED}},\ }\href {https://doi.org/10.1126/sciadv.aaw0297} {\bibfield  {journal} {\bibinfo  {journal} {Sci. Adv.}\ }\textbf {\bibinfo {volume} {5}},\ \bibinfo {pages} {eaaw0297} (\bibinfo {year} {2019})}\BibitemShut {NoStop}%
\bibitem [{\citenamefont {Downing}\ \emph {et~al.}(2019)\citenamefont {Downing}, \citenamefont {Sturges}, \citenamefont {Weick}, \citenamefont {Stobińska},\ and\ \citenamefont {Martín-Moreno}}]{Downing2019}%
  \BibitemOpen
  \bibfield  {author} {\bibinfo {author} {\bibfnamefont {C.~A.}\ \bibnamefont {Downing}}, \bibinfo {author} {\bibfnamefont {T.~J.}\ \bibnamefont {Sturges}}, \bibinfo {author} {\bibfnamefont {G.}~\bibnamefont {Weick}}, \bibinfo {author} {\bibfnamefont {M.}~\bibnamefont {Stobińska}},\ and\ \bibinfo {author} {\bibfnamefont {L.}~\bibnamefont {Martín-Moreno}},\ }\bibfield  {title} {\bibinfo {title} {Topological phases of polaritons in a cavity waveguide},\ }\href {https://doi.org/10.1103/physrevlett.123.217401} {\bibfield  {journal} {\bibinfo  {journal} {Phys. Rev. Lett.}\ }\textbf {\bibinfo {volume} {123}},\ \bibinfo {pages} {217401} (\bibinfo {year} {2019})}\BibitemShut {NoStop}%
\bibitem [{\citenamefont {Malki}\ \emph {et~al.}(2019)\citenamefont {Malki}, \citenamefont {M\"uller},\ and\ \citenamefont {Uhrig}}]{Malki2019}%
  \BibitemOpen
  \bibfield  {author} {\bibinfo {author} {\bibfnamefont {M.}~\bibnamefont {Malki}}, \bibinfo {author} {\bibfnamefont {L.}~\bibnamefont {M\"uller}},\ and\ \bibinfo {author} {\bibfnamefont {G.~S.}\ \bibnamefont {Uhrig}},\ }\bibfield  {title} {\bibinfo {title} {{Absence of localized edge modes in spite of a non-trivial Zak phase in ${\textrm{BiCu}}_{2}{\textrm{PO}}_{6}$}},\ }\href {https://doi.org/10.1103/PhysRevResearch.1.033197} {\bibfield  {journal} {\bibinfo  {journal} {Phys. Rev. Res.}\ }\textbf {\bibinfo {volume} {1}},\ \bibinfo {pages} {033197} (\bibinfo {year} {2019})}\BibitemShut {NoStop}%
\bibitem [{\citenamefont {Nie}\ and\ \citenamefont {Liu}(2020)}]{Nie2020}%
  \BibitemOpen
  \bibfield  {author} {\bibinfo {author} {\bibfnamefont {W.}~\bibnamefont {Nie}}\ and\ \bibinfo {author} {\bibfnamefont {Y.}~\bibnamefont {Liu}},\ }\bibfield  {title} {\bibinfo {title} {Bandgap-assisted quantum control of topological edge states in a cavity},\ }\href {https://doi.org/10.1103/PhysRevResearch.2.012076} {\bibfield  {journal} {\bibinfo  {journal} {Phys. Rev. Res.}\ }\textbf {\bibinfo {volume} {2}},\ \bibinfo {pages} {012076} (\bibinfo {year} {2020})}\BibitemShut {NoStop}%
\bibitem [{\citenamefont {Lin}\ \emph {et~al.}(2020)\citenamefont {Lin}, \citenamefont {Kruk}, \citenamefont {Ke}, \citenamefont {Lee},\ and\ \citenamefont {Kivshar}}]{Ling2020}%
  \BibitemOpen
  \bibfield  {author} {\bibinfo {author} {\bibfnamefont {L.}~\bibnamefont {Lin}}, \bibinfo {author} {\bibfnamefont {S.}~\bibnamefont {Kruk}}, \bibinfo {author} {\bibfnamefont {Y.}~\bibnamefont {Ke}}, \bibinfo {author} {\bibfnamefont {C.}~\bibnamefont {Lee}},\ and\ \bibinfo {author} {\bibfnamefont {Y.}~\bibnamefont {Kivshar}},\ }\bibfield  {title} {\bibinfo {title} {Topological states in disordered arrays of dielectric nanoparticles},\ }\href {https://doi.org/10.1103/PhysRevResearch.2.043233} {\bibfield  {journal} {\bibinfo  {journal} {Phys. Rev. Res.}\ }\textbf {\bibinfo {volume} {2}},\ \bibinfo {pages} {043233} (\bibinfo {year} {2020})}\BibitemShut {NoStop}%
\bibitem [{\citenamefont {Hsu}\ and\ \citenamefont {Chen}(2020)}]{Hsu2020}%
  \BibitemOpen
  \bibfield  {author} {\bibinfo {author} {\bibfnamefont {H.-C.}\ \bibnamefont {Hsu}}\ and\ \bibinfo {author} {\bibfnamefont {T.-W.}\ \bibnamefont {Chen}},\ }\bibfield  {title} {\bibinfo {title} {{Topological Anderson insulating phases in the long-range Su-Schrieffer-Heeger model}},\ }\href {https://doi.org/10.1103/PhysRevB.102.205425} {\bibfield  {journal} {\bibinfo  {journal} {Phys. Rev. B}\ }\textbf {\bibinfo {volume} {102}},\ \bibinfo {pages} {205425} (\bibinfo {year} {2020})}\BibitemShut {NoStop}%
\bibitem [{\citenamefont {Nie}\ \emph {et~al.}(2021)\citenamefont {Nie}, \citenamefont {Shi}, \citenamefont {Nori},\ and\ \citenamefont {Liu}}]{Nie2021}%
  \BibitemOpen
  \bibfield  {author} {\bibinfo {author} {\bibfnamefont {W.}~\bibnamefont {Nie}}, \bibinfo {author} {\bibfnamefont {T.}~\bibnamefont {Shi}}, \bibinfo {author} {\bibfnamefont {F.}~\bibnamefont {Nori}},\ and\ \bibinfo {author} {\bibfnamefont {Y.}~\bibnamefont {Liu}},\ }\bibfield  {title} {\bibinfo {title} {Topology-enhanced nonreciprocal scattering and photon absorption in a waveguide},\ }\href {https://doi.org/10.1103/PhysRevApplied.15.044041} {\bibfield  {journal} {\bibinfo  {journal} {Phys. Rev. Appl.}\ }\textbf {\bibinfo {volume} {15}},\ \bibinfo {pages} {044041} (\bibinfo {year} {2021})}\BibitemShut {NoStop}%
\bibitem [{\citenamefont {Jiao}\ \emph {et~al.}(2021)\citenamefont {Jiao}, \citenamefont {Longhi}, \citenamefont {Wang}, \citenamefont {Gao}, \citenamefont {Zhou}, \citenamefont {Wang}, \citenamefont {Fu}, \citenamefont {Wang}, \citenamefont {Ren}, \citenamefont {Qiao},\ and\ \citenamefont {Jin}}]{Jiao2021}%
  \BibitemOpen
  \bibfield  {author} {\bibinfo {author} {\bibfnamefont {Z.-Q.}\ \bibnamefont {Jiao}}, \bibinfo {author} {\bibfnamefont {S.}~\bibnamefont {Longhi}}, \bibinfo {author} {\bibfnamefont {X.-W.}\ \bibnamefont {Wang}}, \bibinfo {author} {\bibfnamefont {J.}~\bibnamefont {Gao}}, \bibinfo {author} {\bibfnamefont {W.-H.}\ \bibnamefont {Zhou}}, \bibinfo {author} {\bibfnamefont {Y.}~\bibnamefont {Wang}}, \bibinfo {author} {\bibfnamefont {Y.-X.}\ \bibnamefont {Fu}}, \bibinfo {author} {\bibfnamefont {L.}~\bibnamefont {Wang}}, \bibinfo {author} {\bibfnamefont {R.-J.}\ \bibnamefont {Ren}}, \bibinfo {author} {\bibfnamefont {L.-F.}\ \bibnamefont {Qiao}},\ and\ \bibinfo {author} {\bibfnamefont {X.-M.}\ \bibnamefont {Jin}},\ }\bibfield  {title} {\bibinfo {title} {Experimentally detecting quantized {Zak} phases without chiral symmetry in photonic lattices},\ }\href {https://doi.org/10.1103/PhysRevLett.127.147401} {\bibfield  {journal} {\bibinfo  {journal} {Phys. Rev. Lett.}\ }\textbf {\bibinfo {volume} {127}},\ \bibinfo {pages}
  {147401} (\bibinfo {year} {2021})}\BibitemShut {NoStop}%
\bibitem [{\citenamefont {Dias}\ and\ \citenamefont {Marques}(2022)}]{Dias2022}%
  \BibitemOpen
  \bibfield  {author} {\bibinfo {author} {\bibfnamefont {R.~G.}\ \bibnamefont {Dias}}\ and\ \bibinfo {author} {\bibfnamefont {A.~M.}\ \bibnamefont {Marques}},\ }\bibfield  {title} {\bibinfo {title} {Long-range hopping and indexing assumption in one-dimensional topological insulators},\ }\href {https://doi.org/10.1103/PhysRevB.105.035102} {\bibfield  {journal} {\bibinfo  {journal} {Phys. Rev. B}\ }\textbf {\bibinfo {volume} {105}},\ \bibinfo {pages} {035102} (\bibinfo {year} {2022})}\BibitemShut {NoStop}%
\bibitem [{\citenamefont {P\'erez-Gonz\'alez}\ \emph {et~al.}(2022)\citenamefont {P\'erez-Gonz\'alez}, \citenamefont {G\'omez-Le\'on},\ and\ \citenamefont {Platero}}]{PerezGonzalez2022}%
  \BibitemOpen
  \bibfield  {author} {\bibinfo {author} {\bibfnamefont {B.}~\bibnamefont {P\'erez-Gonz\'alez}}, \bibinfo {author} {\bibfnamefont {A.}~\bibnamefont {G\'omez-Le\'on}},\ and\ \bibinfo {author} {\bibfnamefont {G.}~\bibnamefont {Platero}},\ }\bibfield  {title} {\bibinfo {title} {Topology detection in cavity {QED}},\ }\href {https://doi.org/10.1039/D2CP01806C} {\bibfield  {journal} {\bibinfo  {journal} {Phys. Chem. Chem. Phys.}\ }\textbf {\bibinfo {volume} {24}},\ \bibinfo {pages} {15860} (\bibinfo {year} {2022})}\BibitemShut {NoStop}%
\bibitem [{\citenamefont {McDonnell}\ and\ \citenamefont {Olmos}(2022)}]{McDonnell2022}%
  \BibitemOpen
  \bibfield  {author} {\bibinfo {author} {\bibfnamefont {C.}~\bibnamefont {McDonnell}}\ and\ \bibinfo {author} {\bibfnamefont {B.}~\bibnamefont {Olmos}},\ }\bibfield  {title} {\bibinfo {title} {Subradiant edge states in an atom chain with waveguide-mediated hopping},\ }\href {https://doi.org/10.22331/q-2022-09-15-805} {\bibfield  {journal} {\bibinfo  {journal} {{Quantum}}\ }\textbf {\bibinfo {volume} {6}},\ \bibinfo {pages} {805} (\bibinfo {year} {2022})}\BibitemShut {NoStop}%
\bibitem [{\citenamefont {Wei}(2022)}]{Wei2022}%
  \BibitemOpen
  \bibfield  {author} {\bibinfo {author} {\bibfnamefont {J.}~\bibnamefont {Wei}},\ }\bibfield  {title} {\bibinfo {title} {{Cavity-controlled exciton transport of the Su-Schrieffer-Heeger chain}},\ }\href {https://doi.org/10.1103/PhysRevA.106.033710} {\bibfield  {journal} {\bibinfo  {journal} {Phys. Rev. A}\ }\textbf {\bibinfo {volume} {106}},\ \bibinfo {pages} {033710} (\bibinfo {year} {2022})}\BibitemShut {NoStop}%
\bibitem [{\citenamefont {Dmytruk}\ and\ \citenamefont {Schir\`o}(2022)}]{Dmytruk2022}%
  \BibitemOpen
  \bibfield  {author} {\bibinfo {author} {\bibfnamefont {O.}~\bibnamefont {Dmytruk}}\ and\ \bibinfo {author} {\bibfnamefont {M.}~\bibnamefont {Schir\`o}},\ }\bibfield  {title} {\bibinfo {title} {{Controlling topological phases of matter with quantum light}},\ }\href {https://doi.org/10.1038/s42005-022-01049-0} {\bibfield  {journal} {\bibinfo  {journal} {Commun. Phys.}\ }\textbf {\bibinfo {volume} {5}},\ \bibinfo {pages} {271} (\bibinfo {year} {2022})}\BibitemShut {NoStop}%
\bibitem [{\citenamefont {Buend\'ia}\ \emph {et~al.}(2023)\citenamefont {Buend\'ia}, \citenamefont {Sánchez-Gil},\ and\ \citenamefont {Giannini}}]{Buendia2023}%
  \BibitemOpen
  \bibfield  {author} {\bibinfo {author} {\bibfnamefont {A.}~\bibnamefont {Buend\'ia}}, \bibinfo {author} {\bibfnamefont {J.~A.}\ \bibnamefont {Sánchez-Gil}},\ and\ \bibinfo {author} {\bibfnamefont {V.}~\bibnamefont {Giannini}},\ }\bibfield  {title} {\bibinfo {title} {Exploiting oriented field projectors to open topological gaps in plasmonic nanoparticle arrays},\ }\href {https://doi.org/10.1021/acsphotonics.2c01526} {\bibfield  {journal} {\bibinfo  {journal} {ACS Photonics}\ }\textbf {\bibinfo {volume} {10}},\ \bibinfo {pages} {464} (\bibinfo {year} {2023})}\BibitemShut {NoStop}%
\bibitem [{\citenamefont {Pirmoradian}\ \emph {et~al.}(2023)\citenamefont {Pirmoradian}, \citenamefont {Miri}, \citenamefont {Zare~Rameshti},\ and\ \citenamefont {Saeidian}}]{Pirmoradian2023}%
  \BibitemOpen
  \bibfield  {author} {\bibinfo {author} {\bibfnamefont {F.}~\bibnamefont {Pirmoradian}}, \bibinfo {author} {\bibfnamefont {M.}~\bibnamefont {Miri}}, \bibinfo {author} {\bibfnamefont {B.}~\bibnamefont {Zare~Rameshti}},\ and\ \bibinfo {author} {\bibfnamefont {S.}~\bibnamefont {Saeidian}},\ }\bibfield  {title} {\bibinfo {title} {Topological magnon modes of a chain of magnetic spheres in a waveguide},\ }\href {https://doi.org/10.1103/PhysRevB.107.064401} {\bibfield  {journal} {\bibinfo  {journal} {Phys. Rev. B}\ }\textbf {\bibinfo {volume} {107}},\ \bibinfo {pages} {064401} (\bibinfo {year} {2023})}\BibitemShut {NoStop}%
\bibitem [{\citenamefont {Kvande}\ \emph {et~al.}(2023)\citenamefont {Kvande}, \citenamefont {Hill},\ and\ \citenamefont {Blume}}]{Kvande2023}%
  \BibitemOpen
  \bibfield  {author} {\bibinfo {author} {\bibfnamefont {C.~I.}\ \bibnamefont {Kvande}}, \bibinfo {author} {\bibfnamefont {D.~B.}\ \bibnamefont {Hill}},\ and\ \bibinfo {author} {\bibfnamefont {D.}~\bibnamefont {Blume}},\ }\bibfield  {title} {\bibinfo {title} {{Finite Su-Schrieffer-Heeger chains coupled to a two-level emitter: Hybridization of edge and emitter states}},\ }\href {https://doi.org/10.1103/PhysRevA.108.023703} {\bibfield  {journal} {\bibinfo  {journal} {Phys. Rev. A}\ }\textbf {\bibinfo {volume} {108}},\ \bibinfo {pages} {023703} (\bibinfo {year} {2023})}\BibitemShut {NoStop}%
\bibitem [{\citenamefont {Tichauer}\ \emph {et~al.}(2021)\citenamefont {Tichauer}, \citenamefont {Feist},\ and\ \citenamefont {Groenhof}}]{Tichauer2021}%
  \BibitemOpen
  \bibfield  {author} {\bibinfo {author} {\bibfnamefont {R.~H.}\ \bibnamefont {Tichauer}}, \bibinfo {author} {\bibfnamefont {J.}~\bibnamefont {Feist}},\ and\ \bibinfo {author} {\bibfnamefont {G.}~\bibnamefont {Groenhof}},\ }\bibfield  {title} {\bibinfo {title} {Multi-scale dynamics simulations of molecular polaritons: The effect of multiple cavity modes on polariton relaxation},\ }\href {https://doi.org/10.1063/5.0037868} {\bibfield  {journal} {\bibinfo  {journal} {J. Chem. Phys.}\ }\textbf {\bibinfo {volume} {154}},\ \bibinfo {pages} {104112} (\bibinfo {year} {2021})}\BibitemShut {NoStop}%
\bibitem [{\citenamefont {Ribeiro}(2022)}]{Ribeiro2022}%
  \BibitemOpen
  \bibfield  {author} {\bibinfo {author} {\bibfnamefont {R.~F.}\ \bibnamefont {Ribeiro}},\ }\bibfield  {title} {\bibinfo {title} {Multimode polariton effects on molecular energy transport and spectral fluctuations},\ }\href {https://doi.org/10.1038/s42004-022-00660-0} {\bibfield  {journal} {\bibinfo  {journal} {Commun. Chem.}\ }\textbf {\bibinfo {volume} {5}},\ \bibinfo {pages} {48} (\bibinfo {year} {2022})}\BibitemShut {NoStop}%
\bibitem [{\citenamefont {Allard}\ and\ \citenamefont {Weick}(2022)}]{Allard2022}%
  \BibitemOpen
  \bibfield  {author} {\bibinfo {author} {\bibfnamefont {T.~F.}\ \bibnamefont {Allard}}\ and\ \bibinfo {author} {\bibfnamefont {G.}~\bibnamefont {Weick}},\ }\bibfield  {title} {\bibinfo {title} {Disorder-enhanced transport in a chain of lossy dipoles strongly coupled to cavity photons},\ }\href {https://doi.org/10.1103/PhysRevB.106.245424} {\bibfield  {journal} {\bibinfo  {journal} {Phys. Rev. B}\ }\textbf {\bibinfo {volume} {106}},\ \bibinfo {pages} {245424} (\bibinfo {year} {2022})}\BibitemShut {NoStop}%
\bibitem [{\citenamefont {Tichauer}\ \emph {et~al.}(2023)\citenamefont {Tichauer}, \citenamefont {Sokolovskii},\ and\ \citenamefont {Groenhof}}]{Tichauer2023}%
  \BibitemOpen
  \bibfield  {author} {\bibinfo {author} {\bibfnamefont {R.~H.}\ \bibnamefont {Tichauer}}, \bibinfo {author} {\bibfnamefont {I.}~\bibnamefont {Sokolovskii}},\ and\ \bibinfo {author} {\bibfnamefont {G.}~\bibnamefont {Groenhof}},\ }\bibfield  {title} {\bibinfo {title} {{Tuning the Coherent Propagation of Organic Exciton-Polaritons through the Cavity Q-factor}},\ }\href {https://doi.org/https://doi.org/10.1002/advs.202302650} {\bibfield  {journal} {\bibinfo  {journal} {Adv. Sci.}\ }\textbf {\bibinfo {volume} {\!}},\ \bibinfo {pages} {2302650} (\bibinfo {year} {2023})}\BibitemShut {NoStop}%
\bibitem [{\citenamefont {Slobozhanyuk}\ \emph {et~al.}(2015)\citenamefont {Slobozhanyuk}, \citenamefont {Poddubny}, \citenamefont {Miroshnichenko}, \citenamefont {Belov},\ and\ \citenamefont {Kivshar}}]{Slobozhanyuk2015}%
  \BibitemOpen
  \bibfield  {author} {\bibinfo {author} {\bibfnamefont {A.~P.}\ \bibnamefont {Slobozhanyuk}}, \bibinfo {author} {\bibfnamefont {A.~N.}\ \bibnamefont {Poddubny}}, \bibinfo {author} {\bibfnamefont {A.~E.}\ \bibnamefont {Miroshnichenko}}, \bibinfo {author} {\bibfnamefont {P.~A.}\ \bibnamefont {Belov}},\ and\ \bibinfo {author} {\bibfnamefont {Y.~S.}\ \bibnamefont {Kivshar}},\ }\bibfield  {title} {\bibinfo {title} {Subwavelength topological edge states in optically resonant dielectric structures},\ }\href {https://doi.org/10.1103/PhysRevLett.114.123901} {\bibfield  {journal} {\bibinfo  {journal} {Phys. Rev. Lett.}\ }\textbf {\bibinfo {volume} {114}},\ \bibinfo {pages} {123901} (\bibinfo {year} {2015})}\BibitemShut {NoStop}%
\bibitem [{\citenamefont {Zhang}\ \emph {et~al.}(2020)\citenamefont {Zhang}, \citenamefont {Wu}, \citenamefont {Shi},\ and\ \citenamefont {Fung}}]{Zhang2020}%
  \BibitemOpen
  \bibfield  {author} {\bibinfo {author} {\bibfnamefont {Y.}~\bibnamefont {Zhang}}, \bibinfo {author} {\bibfnamefont {R.~P.~H.}\ \bibnamefont {Wu}}, \bibinfo {author} {\bibfnamefont {L.}~\bibnamefont {Shi}},\ and\ \bibinfo {author} {\bibfnamefont {K.~H.}\ \bibnamefont {Fung}},\ }\bibfield  {title} {\bibinfo {title} {Second-order topological photonic modes in dipolar arrays},\ }\href {https://doi.org/10.1021/acsphotonics.0c00160} {\bibfield  {journal} {\bibinfo  {journal} {ACS Photonics}\ }\textbf {\bibinfo {volume} {7}},\ \bibinfo {pages} {2002} (\bibinfo {year} {2020})}\BibitemShut {NoStop}%
\bibitem [{\citenamefont {Allard}\ and\ \citenamefont {Weick}(2021)}]{Allard2021}%
  \BibitemOpen
  \bibfield  {author} {\bibinfo {author} {\bibfnamefont {T.~F.}\ \bibnamefont {Allard}}\ and\ \bibinfo {author} {\bibfnamefont {G.}~\bibnamefont {Weick}},\ }\bibfield  {title} {\bibinfo {title} {Quantum theory of plasmon polaritons in chains of metallic nanoparticles: From near- to far-field coupling regime},\ }\href {https://doi.org/10.1103/PhysRevB.104.125434} {\bibfield  {journal} {\bibinfo  {journal} {Phys. Rev. B}\ }\textbf {\bibinfo {volume} {104}},\ \bibinfo {pages} {125434} (\bibinfo {year} {2021})}\BibitemShut {NoStop}%
\bibitem [{\citenamefont {Pirmoradian}\ \emph {et~al.}(2018)\citenamefont {Pirmoradian}, \citenamefont {Zare~Rameshti}, \citenamefont {Miri},\ and\ \citenamefont {Saeidian}}]{Pirmoradian2018}%
  \BibitemOpen
  \bibfield  {author} {\bibinfo {author} {\bibfnamefont {F.}~\bibnamefont {Pirmoradian}}, \bibinfo {author} {\bibfnamefont {B.}~\bibnamefont {Zare~Rameshti}}, \bibinfo {author} {\bibfnamefont {M.}~\bibnamefont {Miri}},\ and\ \bibinfo {author} {\bibfnamefont {S.}~\bibnamefont {Saeidian}},\ }\bibfield  {title} {\bibinfo {title} {Topological magnon modes in a chain of magnetic spheres},\ }\href {https://doi.org/10.1103/PhysRevB.98.224409} {\bibfield  {journal} {\bibinfo  {journal} {Phys. Rev. B}\ }\textbf {\bibinfo {volume} {98}},\ \bibinfo {pages} {224409} (\bibinfo {year} {2018})}\BibitemShut {NoStop}%
\bibitem [{\citenamefont {Zare~Rameshti}\ \emph {et~al.}(2022)\citenamefont {Zare~Rameshti}, \citenamefont {Viola~Kusminskiy}, \citenamefont {Haigh}, \citenamefont {Usami}, \citenamefont {Lachance-Quirion}, \citenamefont {Nakamura}, \citenamefont {Hu}, \citenamefont {Tang}, \citenamefont {Bauer},\ and\ \citenamefont {Blanter}}]{Rameshti2022}%
  \BibitemOpen
  \bibfield  {author} {\bibinfo {author} {\bibfnamefont {B.}~\bibnamefont {Zare~Rameshti}}, \bibinfo {author} {\bibfnamefont {S.}~\bibnamefont {Viola~Kusminskiy}}, \bibinfo {author} {\bibfnamefont {J.~A.}\ \bibnamefont {Haigh}}, \bibinfo {author} {\bibfnamefont {K.}~\bibnamefont {Usami}}, \bibinfo {author} {\bibfnamefont {D.}~\bibnamefont {Lachance-Quirion}}, \bibinfo {author} {\bibfnamefont {Y.}~\bibnamefont {Nakamura}}, \bibinfo {author} {\bibfnamefont {C.-M.}\ \bibnamefont {Hu}}, \bibinfo {author} {\bibfnamefont {H.~X.}\ \bibnamefont {Tang}}, \bibinfo {author} {\bibfnamefont {G.~E.~W.}\ \bibnamefont {Bauer}},\ and\ \bibinfo {author} {\bibfnamefont {Y.~M.}\ \bibnamefont {Blanter}},\ }\bibfield  {title} {\bibinfo {title} {Cavity magnonics},\ }\href {https://doi.org/10.1016/j.physrep.2022.06.001} {\bibfield  {journal} {\bibinfo  {journal} {Phys. Rep.}\ }\textbf {\bibinfo {volume} {979}},\ \bibinfo {pages} {1} (\bibinfo {year} {2022})}\BibitemShut {NoStop}%
\bibitem [{\citenamefont {Mann}\ \emph {et~al.}(2018)\citenamefont {Mann}, \citenamefont {Sturges}, \citenamefont {Weick}, \citenamefont {Barnes},\ and\ \citenamefont {Mariani}}]{Mann2018}%
  \BibitemOpen
  \bibfield  {author} {\bibinfo {author} {\bibfnamefont {C.-R.}\ \bibnamefont {Mann}}, \bibinfo {author} {\bibfnamefont {T.~J.}\ \bibnamefont {Sturges}}, \bibinfo {author} {\bibfnamefont {G.}~\bibnamefont {Weick}}, \bibinfo {author} {\bibfnamefont {W.~L.}\ \bibnamefont {Barnes}},\ and\ \bibinfo {author} {\bibfnamefont {E.}~\bibnamefont {Mariani}},\ }\bibfield  {title} {\bibinfo {title} {{Manipulating type-I and type-II Dirac polaritons in cavity-embedded honeycomb metasurfaces}},\ }\href {https://doi.org/10.1038/s41467-018-03982-7} {\bibfield  {journal} {\bibinfo  {journal} {Nat. Commun.}\ }\textbf {\bibinfo {volume} {9}},\ \bibinfo {pages} {2194} (\bibinfo {year} {2018})}\BibitemShut {NoStop}%
\bibitem [{\citenamefont {Mann}\ \emph {et~al.}(2020)\citenamefont {Mann}, \citenamefont {Horsley},\ and\ \citenamefont {Mariani}}]{Mann2020}%
  \BibitemOpen
  \bibfield  {author} {\bibinfo {author} {\bibfnamefont {C.-R.}\ \bibnamefont {Mann}}, \bibinfo {author} {\bibfnamefont {S.~A.~R.}\ \bibnamefont {Horsley}},\ and\ \bibinfo {author} {\bibfnamefont {E.}~\bibnamefont {Mariani}},\ }\bibfield  {title} {\bibinfo {title} {Tunable pseudo-magnetic fields for polaritons in strained metasurfaces},\ }\href {https://doi.org/10.1038/s41566-020-0688-8} {\bibfield  {journal} {\bibinfo  {journal} {Nat. Photonics}\ }\textbf {\bibinfo {volume} {14}},\ \bibinfo {pages} {669} (\bibinfo {year} {2020})}\BibitemShut {NoStop}%
\bibitem [{\citenamefont {Yuen-Zhou}\ \emph {et~al.}(2016)\citenamefont {Yuen-Zhou}, \citenamefont {Saikin}, \citenamefont {Zhu}, \citenamefont {Onbasli}, \citenamefont {Ross}, \citenamefont {Bulovic},\ and\ \citenamefont {Baldo}}]{Yuen-Zhou2016}%
  \BibitemOpen
  \bibfield  {author} {\bibinfo {author} {\bibfnamefont {J.}~\bibnamefont {Yuen-Zhou}}, \bibinfo {author} {\bibfnamefont {S.~K.}\ \bibnamefont {Saikin}}, \bibinfo {author} {\bibfnamefont {T.}~\bibnamefont {Zhu}}, \bibinfo {author} {\bibfnamefont {M.~C.}\ \bibnamefont {Onbasli}}, \bibinfo {author} {\bibfnamefont {C.~A.}\ \bibnamefont {Ross}}, \bibinfo {author} {\bibfnamefont {V.}~\bibnamefont {Bulovic}},\ and\ \bibinfo {author} {\bibfnamefont {M.~A.}\ \bibnamefont {Baldo}},\ }\bibfield  {title} {\bibinfo {title} {Plexciton {Dirac} points and topological modes},\ }\href {https://doi.org/10.1038/ncomms11783} {\bibfield  {journal} {\bibinfo  {journal} {Nat. Commun.}\ }\textbf {\bibinfo {volume} {7}},\ \bibinfo {pages} {11783} (\bibinfo {year} {2016})}\BibitemShut {NoStop}%
\bibitem [{\citenamefont {Browaeys}\ \emph {et~al.}(2016)\citenamefont {Browaeys}, \citenamefont {Barredo},\ and\ \citenamefont {Lahaye}}]{Browaeys2016}%
  \BibitemOpen
  \bibfield  {author} {\bibinfo {author} {\bibfnamefont {A.}~\bibnamefont {Browaeys}}, \bibinfo {author} {\bibfnamefont {D.}~\bibnamefont {Barredo}},\ and\ \bibinfo {author} {\bibfnamefont {T.}~\bibnamefont {Lahaye}},\ }\bibfield  {title} {\bibinfo {title} {Experimental investigations of dipole-dipole interactions between a few {Rydberg} atoms},\ }\href {https://doi.org/10.1088/0953-4075/49/15/152001} {\bibfield  {journal} {\bibinfo  {journal} {J. Phys. B}\ }\textbf {\bibinfo {volume} {49}},\ \bibinfo {pages} {152001} (\bibinfo {year} {2016})}\BibitemShut {NoStop}%
\bibitem [{\citenamefont {Perczel}\ \emph {et~al.}(2017)\citenamefont {Perczel}, \citenamefont {Borregaard}, \citenamefont {Chang}, \citenamefont {Pichler}, \citenamefont {Yelin}, \citenamefont {Zoller},\ and\ \citenamefont {Lukin}}]{Perczel2017}%
  \BibitemOpen
  \bibfield  {author} {\bibinfo {author} {\bibfnamefont {J.}~\bibnamefont {Perczel}}, \bibinfo {author} {\bibfnamefont {J.}~\bibnamefont {Borregaard}}, \bibinfo {author} {\bibfnamefont {D.~E.}\ \bibnamefont {Chang}}, \bibinfo {author} {\bibfnamefont {H.}~\bibnamefont {Pichler}}, \bibinfo {author} {\bibfnamefont {S.~F.}\ \bibnamefont {Yelin}}, \bibinfo {author} {\bibfnamefont {P.}~\bibnamefont {Zoller}},\ and\ \bibinfo {author} {\bibfnamefont {M.~D.}\ \bibnamefont {Lukin}},\ }\bibfield  {title} {\bibinfo {title} {Topological quantum optics in two-dimensional atomic arrays},\ }\href {https://doi.org/10.1103/PhysRevLett.119.023603} {\bibfield  {journal} {\bibinfo  {journal} {Phys. Rev. Lett.}\ }\textbf {\bibinfo {volume} {119}},\ \bibinfo {pages} {023603} (\bibinfo {year} {2017})}\BibitemShut {NoStop}%
\bibitem [{\citenamefont {Wang}\ and\ \citenamefont {Zhao}(2018{\natexlab{b}})}]{Wang2018a}%
  \BibitemOpen
  \bibfield  {author} {\bibinfo {author} {\bibfnamefont {B.~X.}\ \bibnamefont {Wang}}\ and\ \bibinfo {author} {\bibfnamefont {C.~Y.}\ \bibnamefont {Zhao}},\ }\bibfield  {title} {\bibinfo {title} {Topological photonic states in one-dimensional dimerized ultracold atomic chains},\ }\href {https://doi.org/10.1103/PhysRevA.98.023808} {\bibfield  {journal} {\bibinfo  {journal} {Phys. Rev. A}\ }\textbf {\bibinfo {volume} {98}},\ \bibinfo {pages} {023808} (\bibinfo {year} {2018}{\natexlab{b}})}\BibitemShut {NoStop}%
\bibitem [{\citenamefont {Asenjo-Garcia}\ \emph {et~al.}(2019)\citenamefont {Asenjo-Garcia}, \citenamefont {Kimble},\ and\ \citenamefont {Chang}}]{Asenjo-Garcia2019}%
  \BibitemOpen
  \bibfield  {author} {\bibinfo {author} {\bibfnamefont {A.}~\bibnamefont {Asenjo-Garcia}}, \bibinfo {author} {\bibfnamefont {H.~J.}\ \bibnamefont {Kimble}},\ and\ \bibinfo {author} {\bibfnamefont {D.~E.}\ \bibnamefont {Chang}},\ }\bibfield  {title} {\bibinfo {title} {Optical waveguiding by atomic entanglement in multilevel atom arrays},\ }\href {https://doi.org/10.1073/pnas.1911467116} {\bibfield  {journal} {\bibinfo  {journal} {Proc. Natl. Acad. Sci. U.S.A.}\ }\textbf {\bibinfo {volume} {116}},\ \bibinfo {pages} {25503} (\bibinfo {year} {2019})}\BibitemShut {NoStop}%
\bibitem [{\citenamefont {Downing}\ and\ \citenamefont {Weick}(2017)}]{Downing2017_Topological}%
  \BibitemOpen
  \bibfield  {author} {\bibinfo {author} {\bibfnamefont {C.~A.}\ \bibnamefont {Downing}}\ and\ \bibinfo {author} {\bibfnamefont {G.}~\bibnamefont {Weick}},\ }\bibfield  {title} {\bibinfo {title} {Topological collective plasmons in bipartite chains of metallic nanoparticles},\ }\href {https://doi.org/10.1103/physrevb.95.125426} {\bibfield  {journal} {\bibinfo  {journal} {Phys. Rev. B}\ }\textbf {\bibinfo {volume} {95}},\ \bibinfo {pages} {125426} (\bibinfo {year} {2017})}\BibitemShut {NoStop}%
\bibitem [{\citenamefont {Ryu}\ \emph {et~al.}(2010)\citenamefont {Ryu}, \citenamefont {Schnyder}, \citenamefont {Furusaki},\ and\ \citenamefont {Ludwig}}]{Ryu_2010}%
  \BibitemOpen
  \bibfield  {author} {\bibinfo {author} {\bibfnamefont {S.}~\bibnamefont {Ryu}}, \bibinfo {author} {\bibfnamefont {A.~P.}\ \bibnamefont {Schnyder}}, \bibinfo {author} {\bibfnamefont {A.}~\bibnamefont {Furusaki}},\ and\ \bibinfo {author} {\bibfnamefont {A.~W.~W.}\ \bibnamefont {Ludwig}},\ }\bibfield  {title} {\bibinfo {title} {Topological insulators and superconductors: {Tenfold} way and dimensional hierarchy},\ }\href {https://doi.org/10.1088/1367-2630/12/6/065010} {\bibfield  {journal} {\bibinfo  {journal} {New J. Phys.}\ }\textbf {\bibinfo {volume} {12}},\ \bibinfo {pages} {065010} (\bibinfo {year} {2010})}\BibitemShut {NoStop}%
\bibitem [{\citenamefont {Chen}\ and\ \citenamefont {Chiou}(2020)}]{Chen2020}%
  \BibitemOpen
  \bibfield  {author} {\bibinfo {author} {\bibfnamefont {B.-H.}\ \bibnamefont {Chen}}\ and\ \bibinfo {author} {\bibfnamefont {D.-W.}\ \bibnamefont {Chiou}},\ }\bibfield  {title} {\bibinfo {title} {An elementary rigorous proof of bulk-boundary correspondence in the generalized {Su-Schrieffer-Heeger} model},\ }\href {https://doi.org/https://doi.org/10.1016/j.physleta.2019.126168} {\bibfield  {journal} {\bibinfo  {journal} {Phys. Lett. A}\ }\textbf {\bibinfo {volume} {384}},\ \bibinfo {pages} {126168} (\bibinfo {year} {2020})}\BibitemShut {NoStop}%
\bibitem [{\citenamefont {Hughes}\ \emph {et~al.}(2011)\citenamefont {Hughes}, \citenamefont {Prodan},\ and\ \citenamefont {Bernevig}}]{Hughes2011}%
  \BibitemOpen
  \bibfield  {author} {\bibinfo {author} {\bibfnamefont {T.~L.}\ \bibnamefont {Hughes}}, \bibinfo {author} {\bibfnamefont {E.}~\bibnamefont {Prodan}},\ and\ \bibinfo {author} {\bibfnamefont {B.~A.}\ \bibnamefont {Bernevig}},\ }\bibfield  {title} {\bibinfo {title} {Inversion-symmetric topological insulators},\ }\href {https://doi.org/10.1103/PhysRevB.83.245132} {\bibfield  {journal} {\bibinfo  {journal} {Phys. Rev. B}\ }\textbf {\bibinfo {volume} {83}},\ \bibinfo {pages} {245132} (\bibinfo {year} {2011})}\BibitemShut {NoStop}%
\bibitem [{\citenamefont {van Miert}\ \emph {et~al.}(2016)\citenamefont {van Miert}, \citenamefont {Ortix},\ and\ \citenamefont {Morais~Smith}}]{Miert2016}%
  \BibitemOpen
  \bibfield  {author} {\bibinfo {author} {\bibfnamefont {G.}~\bibnamefont {van Miert}}, \bibinfo {author} {\bibfnamefont {C.}~\bibnamefont {Ortix}},\ and\ \bibinfo {author} {\bibfnamefont {C.}~\bibnamefont {Morais~Smith}},\ }\bibfield  {title} {\bibinfo {title} {Topological origin of edge states in two-dimensional inversion-symmetric insulators and semimetals},\ }\href {https://doi.org/10.1088/2053-1583/4/1/015023} {\bibfield  {journal} {\bibinfo  {journal} {2D Mater.}\ }\textbf {\bibinfo {volume} {4}},\ \bibinfo {pages} {015023} (\bibinfo {year} {2016})}\BibitemShut {NoStop}%
\bibitem [{\citenamefont {Kakazu}\ and\ \citenamefont {Kim}(1994)}]{Kakazu1994}%
  \BibitemOpen
  \bibfield  {author} {\bibinfo {author} {\bibfnamefont {K.}~\bibnamefont {Kakazu}}\ and\ \bibinfo {author} {\bibfnamefont {Y.~S.}\ \bibnamefont {Kim}},\ }\bibfield  {title} {\bibinfo {title} {Quantization of electromagnetic fields in cavities and spontaneous emission},\ }\href {https://doi.org/10.1103/PhysRevA.50.1830} {\bibfield  {journal} {\bibinfo  {journal} {Phys. Rev. A}\ }\textbf {\bibinfo {volume} {50}},\ \bibinfo {pages} {1830} (\bibinfo {year} {1994})}\BibitemShut {NoStop}%
\bibitem [{\citenamefont {Allard}(2023)}]{Allard2023_PhD}%
  \BibitemOpen
  \bibfield  {author} {\bibinfo {author} {\bibfnamefont {T.}~\bibnamefont {Allard}},\ }\emph {\bibinfo {title} {{Disorder and topology in strongly coupled light-matter systems}}},\ \href@noop {} {\bibinfo {type} {{PhD} thesis}},\ \bibinfo  {school} {{Universit{\'e} de Strasbourg}} (\bibinfo {year} {2023})\BibitemShut {NoStop}%
\bibitem [{\citenamefont {Schrieffer}\ and\ \citenamefont {Wolff}(1966)}]{Schrieffer1966}%
  \BibitemOpen
  \bibfield  {author} {\bibinfo {author} {\bibfnamefont {J.~R.}\ \bibnamefont {Schrieffer}}\ and\ \bibinfo {author} {\bibfnamefont {P.~A.}\ \bibnamefont {Wolff}},\ }\bibfield  {title} {\bibinfo {title} {{Relation between the Anderson and Kondo Hamiltonians}},\ }\href {https://doi.org/10.1103/PhysRev.149.491} {\bibfield  {journal} {\bibinfo  {journal} {Phys. Rev.}\ }\textbf {\bibinfo {volume} {149}},\ \bibinfo {pages} {491} (\bibinfo {year} {1966})}\BibitemShut {NoStop}%
\bibitem [{\citenamefont {Delplace}\ \emph {et~al.}(2011)\citenamefont {Delplace}, \citenamefont {Ullmo},\ and\ \citenamefont {Montambaux}}]{Delplace2011}%
  \BibitemOpen
  \bibfield  {author} {\bibinfo {author} {\bibfnamefont {P.}~\bibnamefont {Delplace}}, \bibinfo {author} {\bibfnamefont {D.}~\bibnamefont {Ullmo}},\ and\ \bibinfo {author} {\bibfnamefont {G.}~\bibnamefont {Montambaux}},\ }\bibfield  {title} {\bibinfo {title} {Zak phase and the existence of edge states in graphene},\ }\href {https://doi.org/10.1103/PhysRevB.84.195452} {\bibfield  {journal} {\bibinfo  {journal} {Phys. Rev. B}\ }\textbf {\bibinfo {volume} {84}},\ \bibinfo {pages} {195452} (\bibinfo {year} {2011})}\BibitemShut {NoStop}%
\bibitem [{\citenamefont {Fuchs}\ and\ \citenamefont {Pi\'echon}(2021)}]{Fuchs2021}%
  \BibitemOpen
  \bibfield  {author} {\bibinfo {author} {\bibfnamefont {J.-N.}\ \bibnamefont {Fuchs}}\ and\ \bibinfo {author} {\bibfnamefont {F.}~\bibnamefont {Pi\'echon}},\ }\bibfield  {title} {\bibinfo {title} {Orbital embedding and topology of one-dimensional two-band insulators},\ }\href {https://doi.org/10.1103/PhysRevB.104.235428} {\bibfield  {journal} {\bibinfo  {journal} {Phys. Rev. B}\ }\textbf {\bibinfo {volume} {104}},\ \bibinfo {pages} {235428} (\bibinfo {year} {2021})}\BibitemShut {NoStop}%
\bibitem [{\citenamefont {Wang}\ \emph {et~al.}(2019)\citenamefont {Wang}, \citenamefont {Guo},\ and\ \citenamefont {Jiang}}]{Wang_2019}%
  \BibitemOpen
  \bibfield  {author} {\bibinfo {author} {\bibfnamefont {H.-X.}\ \bibnamefont {Wang}}, \bibinfo {author} {\bibfnamefont {G.-Y.}\ \bibnamefont {Guo}},\ and\ \bibinfo {author} {\bibfnamefont {J.-H.}\ \bibnamefont {Jiang}},\ }\bibfield  {title} {\bibinfo {title} {Band topology in classical waves: Wilson-loop approach to topological numbers and fragile topology},\ }\href {https://doi.org/10.1088/1367-2630/ab3f71} {\bibfield  {journal} {\bibinfo  {journal} {New J. Phys.}\ }\textbf {\bibinfo {volume} {21}},\ \bibinfo {pages} {093029} (\bibinfo {year} {2019})}\BibitemShut {NoStop}%
\bibitem [{\citenamefont {Chang}\ \emph {et~al.}()\citenamefont {Chang}, \citenamefont {Torres}, \citenamefont {Manrique}, \citenamefont {Robles}, \citenamefont {Silalahi}, \citenamefont {Wu}, \citenamefont {Wang}, \citenamefont {Marcucci}, \citenamefont {Pilozzi}, \citenamefont {Conti}, \citenamefont {Lee},\ and\ \citenamefont {Kuo}}]{Chang2022_arXiv}%
  \BibitemOpen
  \bibfield  {author} {\bibinfo {author} {\bibfnamefont {Y.}~\bibnamefont {Chang}}, \bibinfo {author} {\bibfnamefont {N.~D.~R.}\ \bibnamefont {Torres}}, \bibinfo {author} {\bibfnamefont {S.~F.}\ \bibnamefont {Manrique}}, \bibinfo {author} {\bibfnamefont {R.~A.~R.}\ \bibnamefont {Robles}}, \bibinfo {author} {\bibfnamefont {V.~C.}\ \bibnamefont {Silalahi}}, \bibinfo {author} {\bibfnamefont {C.}~\bibnamefont {Wu}}, \bibinfo {author} {\bibfnamefont {G.}~\bibnamefont {Wang}}, \bibinfo {author} {\bibfnamefont {G.}~\bibnamefont {Marcucci}}, \bibinfo {author} {\bibfnamefont {L.}~\bibnamefont {Pilozzi}}, \bibinfo {author} {\bibfnamefont {C.}~\bibnamefont {Conti}}, \bibinfo {author} {\bibfnamefont {R.}~\bibnamefont {Lee}},\ and\ \bibinfo {author} {\bibfnamefont {W.}~\bibnamefont {Kuo}},\ }\bibfield  {title} {\bibinfo {title} {{Probing topological protected transport in finite-sized Su-Schrieffer-Heeger chains}},\ }\Eprint {https://arxiv.org/abs/arXiv:2004.09282} {arXiv:2004.09282} \BibitemShut {NoStop}%
\bibitem [{\citenamefont {Jackson}(2007)}]{Jackson2007}%
  \BibitemOpen
  \bibfield  {author} {\bibinfo {author} {\bibfnamefont {J.~D.}\ \bibnamefont {Jackson}},\ }\href@noop {} {\emph {\bibinfo {title} {Classical Electrodynamics}}},\ \bibinfo {edition} {3rd}\ ed.\ (\bibinfo  {publisher} {Wiley},\ \bibinfo {address} {New York},\ \bibinfo {year} {2007})\BibitemShut {NoStop}%
\bibitem [{\citenamefont {Brandstetter-Kunc}\ \emph {et~al.}(2016)\citenamefont {Brandstetter-Kunc}, \citenamefont {Weick}, \citenamefont {Downing}, \citenamefont {Weinmann},\ and\ \citenamefont {Jalabert}}]{BrandstetterKunc2016}%
  \BibitemOpen
  \bibfield  {author} {\bibinfo {author} {\bibfnamefont {A.}~\bibnamefont {Brandstetter-Kunc}}, \bibinfo {author} {\bibfnamefont {G.}~\bibnamefont {Weick}}, \bibinfo {author} {\bibfnamefont {C.~A.}\ \bibnamefont {Downing}}, \bibinfo {author} {\bibfnamefont {D.}~\bibnamefont {Weinmann}},\ and\ \bibinfo {author} {\bibfnamefont {R.~A.}\ \bibnamefont {Jalabert}},\ }\bibfield  {title} {\bibinfo {title} {Nonradiative limitations to plasmon propagation in chains of metallic nanoparticles},\ }\href {https://doi.org/10.1103/physrevb.94.205432} {\bibfield  {journal} {\bibinfo  {journal} {Phys. Rev. B}\ }\textbf {\bibinfo {volume} {94}},\ \bibinfo {pages} {205432} (\bibinfo {year} {2016})}\BibitemShut {NoStop}%
\bibitem [{\citenamefont {Houdr\'e}\ \emph {et~al.}(1996)\citenamefont {Houdr\'e}, \citenamefont {Stanley},\ and\ \citenamefont {Ilegems}}]{Houdre1996}%
  \BibitemOpen
  \bibfield  {author} {\bibinfo {author} {\bibfnamefont {R.}~\bibnamefont {Houdr\'e}}, \bibinfo {author} {\bibfnamefont {R.~P.}\ \bibnamefont {Stanley}},\ and\ \bibinfo {author} {\bibfnamefont {M.}~\bibnamefont {Ilegems}},\ }\bibfield  {title} {\bibinfo {title} {{Vacuum-field Rabi splitting in the presence of inhomogeneous broadening: Resolution of a homogeneous linewidth in an inhomogeneously broadened system}},\ }\href {https://doi.org/10.1103/PhysRevA.53.2711} {\bibfield  {journal} {\bibinfo  {journal} {Phys. Rev. A}\ }\textbf {\bibinfo {volume} {53}},\ \bibinfo {pages} {2711} (\bibinfo {year} {1996})}\BibitemShut {NoStop}%
\bibitem [{\citenamefont {Michetti}\ and\ \citenamefont {La~Rocca}(2005)}]{Michetti_2005}%
  \BibitemOpen
  \bibfield  {author} {\bibinfo {author} {\bibfnamefont {P.}~\bibnamefont {Michetti}}\ and\ \bibinfo {author} {\bibfnamefont {G.~C.}\ \bibnamefont {La~Rocca}},\ }\bibfield  {title} {\bibinfo {title} {Polariton states in disordered organic microcavities},\ }\href {https://doi.org/10.1103/PhysRevB.71.115320} {\bibfield  {journal} {\bibinfo  {journal} {Phys. Rev. B}\ }\textbf {\bibinfo {volume} {71}},\ \bibinfo {pages} {115320} (\bibinfo {year} {2005})}\BibitemShut {NoStop}%
\bibitem [{\citenamefont {Kockum}\ \emph {et~al.}(2019)\citenamefont {Kockum}, \citenamefont {Miranowicz}, \citenamefont {{De Liberato}}, \citenamefont {Savasta},\ and\ \citenamefont {Nori}}]{Kockum2019_review}%
  \BibitemOpen
  \bibfield  {author} {\bibinfo {author} {\bibfnamefont {A.~F.}\ \bibnamefont {Kockum}}, \bibinfo {author} {\bibfnamefont {A.}~\bibnamefont {Miranowicz}}, \bibinfo {author} {\bibfnamefont {S.}~\bibnamefont {{De Liberato}}}, \bibinfo {author} {\bibfnamefont {S.}~\bibnamefont {Savasta}},\ and\ \bibinfo {author} {\bibfnamefont {F.}~\bibnamefont {Nori}},\ }\bibfield  {title} {\bibinfo {title} {Ultrastrong coupling between light and matter},\ }\href {https://doi.org/10.1038/s42254-018-0006-2} {\bibfield  {journal} {\bibinfo  {journal} {Nat. Rev. Phys.}\ }\textbf {\bibinfo {volume} {1}},\ \bibinfo {pages} {19} (\bibinfo {year} {2019})}\BibitemShut {NoStop}%
\bibitem [{\citenamefont {Forn-Díaz}\ \emph {et~al.}(2019)\citenamefont {Forn-Díaz}, \citenamefont {Lamata}, \citenamefont {Rico}, \citenamefont {Kono},\ and\ \citenamefont {Solano}}]{FornDiaz2019}%
  \BibitemOpen
  \bibfield  {author} {\bibinfo {author} {\bibfnamefont {P.}~\bibnamefont {Forn-Díaz}}, \bibinfo {author} {\bibfnamefont {L.}~\bibnamefont {Lamata}}, \bibinfo {author} {\bibfnamefont {E.}~\bibnamefont {Rico}}, \bibinfo {author} {\bibfnamefont {J.}~\bibnamefont {Kono}},\ and\ \bibinfo {author} {\bibfnamefont {E.}~\bibnamefont {Solano}},\ }\bibfield  {title} {\bibinfo {title} {Ultrastrong coupling regimes of light-matter interaction},\ }\href {https://doi.org/10.1103/RevModPhys.91.025005} {\bibfield  {journal} {\bibinfo  {journal} {Rev. Mod. Phys.}\ }\textbf {\bibinfo {volume} {91}},\ \bibinfo {pages} {025005} (\bibinfo {year} {2019})}\BibitemShut {NoStop}%
\bibitem [{\citenamefont {Masuki}\ and\ \citenamefont {Ashida}(2023)}]{Masuki_arxiv_2022}%
  \BibitemOpen
  \bibfield  {author} {\bibinfo {author} {\bibfnamefont {K.}~\bibnamefont {Masuki}}\ and\ \bibinfo {author} {\bibfnamefont {Y.}~\bibnamefont {Ashida}},\ }\bibfield  {title} {\bibinfo {title} {Berry phase and topology in ultrastrongly coupled quantum light-matter systems},\ }\href {https://doi.org/10.1103/PhysRevB.107.195104} {\bibfield  {journal} {\bibinfo  {journal} {Phys. Rev. B}\ }\textbf {\bibinfo {volume} {107}},\ \bibinfo {pages} {195104} (\bibinfo {year} {2023})}\BibitemShut {NoStop}%
\bibitem [{\citenamefont {Craig}\ and\ \citenamefont {Thirunamachandran}(1984)}]{Craig2012}%
  \BibitemOpen
  \bibfield  {author} {\bibinfo {author} {\bibfnamefont {D.~P.}\ \bibnamefont {Craig}}\ and\ \bibinfo {author} {\bibfnamefont {T.}~\bibnamefont {Thirunamachandran}},\ }\href@noop {} {\emph {\bibinfo {title} {Molecular {Quantum} {Electrodynamics}}}}\ (\bibinfo  {publisher} {Academic Press},\ \bibinfo {address} {London},\ \bibinfo {year} {1984})\BibitemShut {NoStop}%
\bibitem [{\citenamefont {Downing}\ and\ \citenamefont {Martín-Moreno}(2021)}]{Downing2021}%
  \BibitemOpen
  \bibfield  {author} {\bibinfo {author} {\bibfnamefont {C.~A.}\ \bibnamefont {Downing}}\ and\ \bibinfo {author} {\bibfnamefont {L.}~\bibnamefont {Martín-Moreno}},\ }\bibfield  {title} {\bibinfo {title} {{Polaritonic Tamm states induced by cavity photons}},\ }\href {https://doi.org/https://doi.org/10.1515/nanoph-2020-0370} {\bibfield  {journal} {\bibinfo  {journal} {Nanophotonics}\ }\textbf {\bibinfo {volume} {10}},\ \bibinfo {pages} {513 } (\bibinfo {year} {2021})}\BibitemShut {NoStop}%
\bibitem [{\citenamefont {Stefano}\ \emph {et~al.}(2019)\citenamefont {Stefano}, \citenamefont {Settineri}, \citenamefont {Macrì}, \citenamefont {Garziano}, \citenamefont {Stassi}, \citenamefont {Savasta},\ and\ \citenamefont {Nori}}]{Stefano2019}%
  \BibitemOpen
  \bibfield  {author} {\bibinfo {author} {\bibfnamefont {O.~D.}\ \bibnamefont {Stefano}}, \bibinfo {author} {\bibfnamefont {A.}~\bibnamefont {Settineri}}, \bibinfo {author} {\bibfnamefont {V.}~\bibnamefont {Macrì}}, \bibinfo {author} {\bibfnamefont {L.}~\bibnamefont {Garziano}}, \bibinfo {author} {\bibfnamefont {R.}~\bibnamefont {Stassi}}, \bibinfo {author} {\bibfnamefont {S.}~\bibnamefont {Savasta}},\ and\ \bibinfo {author} {\bibfnamefont {F.}~\bibnamefont {Nori}},\ }\bibfield  {title} {\bibinfo {title} {Resolution of gauge ambiguities in ultrastrong-coupling cavity quantum electrodynamics},\ }\href {https://doi.org/10.1038/s41567-019-0534-4} {\bibfield  {journal} {\bibinfo  {journal} {Nat. Phys.}\ }\textbf {\bibinfo {volume} {15}},\ \bibinfo {pages} {803} (\bibinfo {year} {2019})}\BibitemShut {NoStop}%
\end{thebibliography}%
\end{document}